\documentclass[10pt,twocolumn, journal]{IEEEtran}
\usepackage{amsfonts}
\usepackage{dsfont}
\usepackage{setspace}

\usepackage{setspace}
\usepackage{amsmath}
\usepackage{amssymb}
\usepackage[dvips]{graphicx}
\usepackage{epsfig}
\usepackage{dsfont}
\usepackage[ps2pdf,
bookmarks=false,
bookmarksnumbered=false, 
bookmarksopen=false, 
colorlinks=false]{}

\newcommand{\tsnr}{{\text{\footnotesize{SNR}}}}

\newcommand{\E}{\mathbb{E}}

\newcommand{\Pb}{\bar{P}}

\newcommand{\figsize}{0.45}

\newcommand{\R}{{\mathbf{R}}}
\newcommand{\Or}{{\boldsymbol{\pi}}}
\newcommand{\bD}{{\mathbf{D}}}
\newcommand{\bS}{{\mathbf{S}}}
\newcommand{\aR}{{\mathrm{R}}}

\newcommand{\pra}{2^{\frac{r_{\bS_1,\R}}{\tau B}}}
\newcommand{\prb}{2^{\frac{r_{\bS_2,\R}}{\tau B}}}
\newcommand{\prc}{2^{\frac{r_{\R,\bD_1}}{(1-\tau) B}}}
\newcommand{\prd}{2^{\frac{r_{\R,\bD_2}}{(1-\tau) B}}}

\newtheorem{Lem}{Theorem}

\begin{document}

%
\title{On the Throughput of Multi-Source Multi-Destination Relay Networks with Queueing Constraints}

\author{Yi Li, M. Cenk Gursoy and Senem
Velipasalar
\thanks{The authors are with the Department of Electrical
Engineering and Computer Science, Syracuse University, Syracuse, NY, 13244
(e-mail: yli33@syr.edu, mcgursoy@syr.edu, svelipas@syr.edu).}
\thanks{The material in this paper was presented in part at the IEEE Wireless Communications and Networking Conference (WCNC), New Orleans, LA, March 2015.}}

\maketitle


\begin{abstract}
In this paper, the throughput of relay networks with multiple source-destination pairs under queueing constraints has been investigated for both variable-rate and fixed-rate schemes. When channel side information (CSI) is available at the transmitter side, transmitters can adapt their transmission rates according to the channel conditions, and achieve the instantaneous channel capacities. In this case, the departure rates at each node have been characterized for different system parameters, which control the power allocation, time allocation and decoding order. In the other case of no CSI at the transmitters, a simple automatic repeat request (ARQ) protocol with fixed rate transmission is used to provide reliable communication. Under this ARQ assumption, the instantaneous departure rates at each node can be modeled as an ON-OFF process, and the probabilities of ON and OFF states are identified. With the characterization of the arrival and departure rates at each buffer, stability conditions are identified and effective capacity analysis is conducted for both cases to determine the system throughput under statistical queueing constraints. In addition, for the variable-rate scheme, the concavity of the sum rate is shown for certain parameters, helping to improve the efficiency of parameter optimization. Finally, via numerical results, the influence of system parameters and the behavior of the system throughput are identified.
\end{abstract}

\begin{IEEEkeywords}
broadcast channel, buffer overflow, decode-and-forward relaying, effective capacity, fixed-rate transmissions,  multiple-access channel, statistical queueing constraints, throughput, variable-rate transmissions.
\end{IEEEkeywords}

\thispagestyle{empty}

\section{Introduction}
Increasing transmission rates, improving energy efficiency and enhancing reliability are important considerations in wireless communications. Various advanced schemes have been proposed  to address these concerns. One such strategy is cooperative communications. In particular, relay networks can greatly enhance the performance for long distance transmissions among users and improve resource efficiency. For instance, the throughput of relay networks have been analyzed by several studies. In \cite{Gaussian_OMARC}, the achievable rates of Gaussian orthogonal multi-access relay channels in which multiple sources communicate with one destination with the help of one relay were investigated, which were also proved to have a max-flow min-cut interpretation. The throughput region of the same system model was also given in \cite{Region_OMARC} with superposition block Markov encoding and multiple access encoding. Further analysis was also provided in \cite{OMARC_optimal}, in which the optimal resource allocation strategy was studied to achieve the maximum sum rate. In \cite{OMARN}, the system throughput region of the generalized multiple access relay network , which includes multiple transmitters, multiple relays, and a single destination, was studied. In all cases, with the help of relay nodes, the channel conditions effectively improve for long distance wireless communication, and performance enhancements are realized.

A further generalization of multiple-access relay channels is to introduce multiple destination nodes. These models are referred to as multi-source multi-destination relay networks. Multi-source multi-destination relay network model can be seen as a combination of multiple-access, broadcast, and two-hop relay channels, and it can be used to address scenarios in which multiple pairs of users simultaneously communicate with the help of a relay node. A basic practical example of these models is cellular operation in which multiple mobile users within a cell communicate with each other through a base station, which essentially acts as a relay unit between the source and destination nodes\footnote{Moreover, in LTE-Advanced cellular standards, relaying and coordinated multi point (CoMP) operation  are introduced to provide enhanced coverage and capacity at cell edges, and multi-user relay models can be realized in these operation modes as well.}.  Such networks have been analyzed in several recent studies. In \cite{AF_sumrate}, the throughput of the amplify-and-forward multi-source multi-destination relay network was studied, when the relay was equipped with multiple antennas. Based on this work, the same authors studied the impact of imperfect CSI in \cite{AF_sumrate2}, and proposed an antenna selection algorithm to improve the performance. In \cite{jointpower_opt}, the joint power optimization was investigated for the multi-source multi-destination relay network, and in \cite{MSMDR_NC}, network coding was applied to this type of network, and the system performance was evaluated.

In addition to cooperative operation, due to the critical delay/buffer requirements in real-time data transmissions, such as in live video streaming, quality of service (QoS) guarantees should be provided for acceptable performance in wireless systems supporting multimedia traffic. With this motivation, we consider the throughput of the multi-source multi-destination relay networks under statistical queueing constraints, imposed as limitations on the decay rate of buffer overflow probabilities at all nodes in the system. In \cite{chang1}, effective bandwidth was introduced as a measure of the system throughput under such statistical queueing or QoS constraints. More specifically, effective bandwidth has been defined as the minimum constant transmission rate required to support time-varying arrivals while the buffer overflow probability decays exponentially with increasing overflow threshold. In \cite{chang2}, effective bandwidths of departure processes with time-varying service rates were investigated, and the theory of effective bandwidth was employed to analyze the performance of high speed networks in \cite{chang3}. Later, effective capacity was defined in \cite{dapeng} as a dual concept to characterize the maximum constant arrival rates that can be supported by time-varying wireless transmission rates again under statistical queueing constraints. 

Recently, effective capacity analysis has been applied to multiuser and cooperative relay systems. For cooperative relay systems, the authors in \cite{tangrelay} studied efficient resource allocation strategies over wireless relay channels under statistical QoS constraints by employing effective capacity as the throughput metric. However, in this work, either no buffer was needed at the relay if amplify-and-forward (AF) strategy was employed or no relay buffer constraints were imposed when decode-and-forward (DF) was used. In \cite{butterfly}, queueing analysis was conducted for a butterfly network when the arrivals were modeled as a two-state Markov-modulated fluid process, and network coding or classical routing was performed by the intermediate relay node. In this study, all links were assumed to be time-invariant. Therefore, static rather than fading channels were considered. In \cite{deli-relay}, effective capacity of a two-hop wireless link was determined, characterizing the maximum constant arrival rates in such systems under queueing constraints at both the source and relay nodes. This two-hop model consisted of only one source and one destination, and consequently there were no multiple-access and broadcast phases unlike in our multi-source multi-destination model, and no considerations regarding rate selection from multiple-user achievable rate regions, interference from concurrent transmissions, and decoding order selection at the receivers.
In \cite{Deli_MAC}, the effective capacity region of the multiple-access fading channel under queueing constraints was analyzed, and this result was extended to characterize the throughput region of the multiple-access channel with Markov arrivals in \cite{Markov_MAC}. The effective capacity region of the fading broadcast channel and optimal power allocation policies were studied in \cite{Deli_BC}. These studies addressed single-hop channels and  did not consider cooperative schemes. Therefore, with respect to all above-mentioned related prior studies, the key novelty in this paper is the throughput analysis under statistical queueing constraints at all transmitting nodes for a two-hop multi-source multi-destination cooperative network model which combines multiple-access, broadcast, and relay fading channels.


As noted before, beside QoS requirements, reliability and robustness are important concerns in wireless systems, especially when CSI is not available at the transmitter. In this situation, data transmission can be performed at fixed rates, and reliability can be ensured via automatic repeat request (ARQ) protocols, which trigger retransmissions in cases of decoding failure. This effectively enables the transmitter to adapt to the channel conditions with only limited feedback from the receiver. Queueing analysis has also been performed when ARQ is employed in the communication system. For instance, in \cite{queue_arq}, queueing models were formulated and performance analysis was conducted for go-back-N and selective repeat ARQ protocols, and the energy efficiency of ARQ with fixed transmission rates was analyzed in \cite{ARQ_fix} under statistical queueing constraints.

In this paper, the system throughput of multi-source multi-destination relay network is investigated under statistical queueing constraints primarily for a network model with two source-destination pairs and one intermediate relay. The following are our main contributions:
\begin{enumerate}
  \item We characterize the throughput of the multi-source multi-destination relay network under queueing constraints by using the stochastic network calculus framework and effective capacity formulations. We identify the impact of resource allocation policies and decoding strategies on the performance.
  \item We extend our analysis to the network with more than two source-destination pairs and also to the model with full-duplex relay operation.
  \item We perform an effective capacity analysis for the case in which CSI is not available at the transmitter nodes, all transmitters are sending the data at fixed rates, and an ARQ protocol is employed.
\end{enumerate}

The rest of this paper is organized as follows. We describe our system model in Section \ref{sec:system-model}, and introduce preliminary concepts regarding statistical queueing constraints and arrival rates in Section \ref{sec_Prel}. In Section IV, system throughput is characterized when CSI is available at the transmitters, and several properties of the throughput are identified. In Section V, extensions of the system model treated in Section III are addressed. In Section VI, we analyze the system throughput when there is no CSI at the transmitters. Finally, we draw conclusions in Section VII.

\section{System Model} \label{sec:system-model}

\begin{figure}
\begin{center}
\includegraphics[width=\figsize\textwidth]{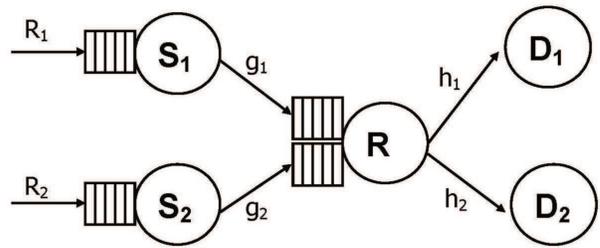}
\caption{The relay network system with buffer constraints.}\label{fig:systemmodel}
\end{center}
\end{figure}

In this paper, we consider a multi-source multi-destination relay network model with two pairs of sources and destinations, as depicted in Figure \ref{fig:systemmodel}. In this system, two sources $\bS_1$ and $\bS_2$ send information to their corresponding destinations $\bD_1$ and $\bD_2$ with the help of an intermediate relay node, and there is no direct link between the source nodes and their destinations. This assumption is accurate if the source and destination nodes are sufficiently far apart in distance. We assume that $\bD_j$ only needs the packets coming from source $\bS_j$, where $j=1,2$. Each source node has a buffer, keeping the packets to be transmitted to the relay node. The arrival rates at source nodes $\bS_1$ and $\bS_2$ are assumed to be constant, and are denoted as $\aR_1$ and $\aR_2$ respectively.  At the relay node, there are two buffers\footnote{In practice, only one physical buffer is sufficient at the relay node to store the received packets from $\bS_1$ and $\bS_2$. In the analysis, we essentially decompose this physical buffer into two equivalent virtual buffers, in each of which data for only one destination is stored and first-in first-out policy is employed.}, one for keeping the decoded information coming from source $\bS_1$, and the other for the decoded data of $\bS_2$.

In our setup, relay node performs decode-and-forward relaying and works in half-duplex mode, and hence it cannot transmit and receive at the same time. The entire transmission process can be divided into two phases, namely multiple-access phase and broadcast phase. In the multiple-access phase, both $\bS_1$ and $\bS_2$ transmit to the relay node simultaneously through a multiple-access channel. Relay node attempts to decode their messages by using certain decoding orders, and the decoded information bits are stored in their corresponding buffers at the relay. We assume that if fixed-rate transmissions are employed, transmission fails if the rate is greater than the instantaneous capacity of the link for a given decoding strategy at the relay\footnote{It is  assumed that errors are detected reliably at the receivers, and when the system employs ARQ protocol, acknowledgement (ACK) and retransmission request (RQ) packets are assumed to be received with no errors.}.

The received discrete-time signal at the relay node can be expressed as
\begin{align}\label{eq:RSG_R}
Y_r[i]&=g_1[i]X_1[i]+g_2[i]X_2[i]+n_r[i],
\end{align}
where $X_j$ for $j = 1,2$ represents the transmitted signal from source node $\bS_j$, $g_j$ is the fading coefficient of the $\bS_j-\R$ link, and $n_r$ is the additive Gaussian noise at the relay.

In the broadcast phase, relay node forwards information bits to their destinations through a broadcast channel. The received signal at $\bD_j$ is
\begin{align}
Y_j[i]&=h_j[i]X_r[i]+n_j[i], \quad j = 1, 2
\end{align}
where $X_r$ stands for the transmitted signal from $\R$, $n_j$ is the additive Gaussian noise at $\bD_j$, and $h_j$ represents the channel fading coefficient of the $\R-\bD_j$ link. Magnitude-squares of the fading coefficients in both phases are denoted by $z_j[i]=|g_j[i]|^2$ and $\omega_j[i]=|h_j[i]|^2$, for $j=1,2$. In our analysis, we consider block fading and assume that fading coefficients stay constant in one time block, and change independently from block to block. While our analysis is general and applicable to any fading distribution with finite variances, we assume Rayleigh fading in all channels in our numerical analysis.

The transmitted signals are subject to energy constraints given by $\E\{|X_j|^2\}\le \Pb_j/B$ for $j=1,2$ and $\E\{|X_r|^2\}\le\Pb_r/B$, where $B$ is the system bandwidth and $\Pb_k$ for $k=1,2,r$ is the transmit power constraint for the corresponding node. The additive noise terms $n_k[i]$ for $k=1,2,r$ are independent, zero-mean, circularly symmetric, complex Gaussian random variables with variances
$\E\{|n_k[i]|^2\} = N_0$. Then, signal-to-noise ratios are defined as
\begin{equation}
\tsnr_k=\frac{\Pb_k}{N_0 B}
\end{equation}
where $k=1,2,r$.

Finally, there are three important system parameters: $\tau$, $\rho$ and $\delta$. $\tau \in (0,1)$ denotes the fraction of time allocated to the multiple-access phase, and hence the fraction of time allocated to the broadcast phase is $1-\tau$. $\rho \in (0,1)$ represents the fraction of power allocated by the relay to the transmission of the message intended for $\bD_1$, and therefore the fraction of power allocated to the transmission to $\bD_2$ is $1-\rho$. In the multiple-access phase, relay node decodes the received signal using different decoding orders, and the fraction of time allocated to decoding order $\{1,2\}$ and $\{2,1\}$ at the relay node are denoted by $\delta$ and $1-\delta$, respectively. This time sharing strategy between different decoding orders is used only for the case of variable-rate transmissions, performed when CSI is available at all transmitters. For fixed-rate transmission schemes, decoding order is part of the decoding strategy, which is fixed for each node.

\section{Preliminaries on Statistical Queueing Constraints and Arrival Rates}\label{sec_Prel}
In our work, we assume that the queueing constraints are imposed so that buffer overflow probability decays exponentially fast, i.e., we have
\begin{gather}
\label{buffer_vio1}
\Pr\{Q \ge q_{\max}\} \approx \gamma e^{-\theta q_{\max}}
\end{gather}
where $Q$ is the stationary queue length, $q_{\max}$ is a sufficiently large buffer overflow threshold, $\gamma=\Pr\{Q>0\}$ is the probability that buffer is non-empty, and $\theta$ is called the QoS exponent. This QoS exponent can more precisely be formulated as
\begin{gather}
\theta = \lim_{q_{\max} \to \infty} \frac{-\log \Pr\{Q \ge q_{\max}\}}{q_{\max}}.
\end{gather}
It is obvious that a larger $\theta$ value implies stricter constraints on the buffer overflows.

In our analysis, the departure process from each buffer is assumed to be a stationary process. We first define the asymptotic logarithmic moment generating function (LMGF) of an arrival or service process $a[i]$ as a function of the QoS parameter $\theta$ as \footnote{Throughout the text,
logarithm expressed without a base, i.e., $\log(\cdot)$, refers to the natural logarithm $\log_e(\cdot)$.}
\begin{equation}\label{LMGF_df}
\Lambda_A(\theta)=\lim_{n\to\infty}\frac{\log \E\{e^{\theta\sum_{i=1}^n
a[i]}\}}{n}.
\end{equation}
It can be easily verified that the asymptotic LMGF of a constant-rate arrival process, $a[i]=R$, is $\theta R$.

In our multi-source multi-destination relay system, we assume that the QoS exponents at the source nodes $\bS_1$ and $\bS_2$ and the relay node $\R$ are denoted by $\theta_1$, $\theta_2$ and $\theta_r$, respectively. Hence, we wish to have the buffer overflow probability at node $j \in \{1,2,r\}$ to behave approximately as $\Pr\{Q_j \ge q_{\max}\} \approx \gamma e^{-\theta_j q_{\max}}$ for large $q_{\max}$.

From the theory of effective bandwidth and effective capacity \cite{chang1}, \cite{chang2}, \cite{dapeng}, the buffer overflow probability decays exponentially as $e^{-\theta_j q_{\max}}$ or faster at $\bS_j$ if the constant arrival rate at $\bS_j$ satisfies
\begin{align}\label{eq:cond1equ}
\aR_j=-\frac{\Lambda_{\bS_j,\R}(-\tilde{\theta}_j)}{\tilde{\theta}_j},\,j=1,2
\end{align}
for some $\tilde{\theta}_j \ge \theta_j$. Above $\Lambda_{\bS_j,\R}$ is the asymptotic LMGF of service process at $\bS_j$.

The above arrival rate formulation considers only the queueing constraints at the source nodes. However, we need to address the constraints at the relay buffers as well. With the characterization of the effective bandwidth of departure processes in queues with time-varying service rates, it was shown in \cite{chang2} that the buffer overflow probabilities at the relay decay as $e^{-\theta_r q_{\max}}$ or faster for large $q_{\max}$ if we have
\begin{align}\label{eq:cond2equ1}
\Lambda_{\R}(\hat{\theta}_j)+\Lambda_{\R,\bD_j}(-\hat{\theta}_j)=0
\end{align}
for some $\hat{\theta}_j\ge\theta_r$, $j=1,2$. Above, $\Lambda_{\R,\bD_j}$ is the LMGF of the service rate at $\R$ for the transmission of the message to $\bD_j$.

In (\ref{eq:cond2equ1}), $\Lambda_{\R}$ is the asymptotic LGMF of the arrival process to $\R$ (or equivalently the departure process from $\bS_j$) and is formulated as \cite[equation (18)]{chang2}
\begin{align}
\Lambda_{\R}(\theta)=\left\{\begin{array}{ll}\aR_j\theta, &
0\le\theta\le\tilde{\theta}_j\\\aR_j\tilde{\theta}_j+\Lambda_{\bS_j,\R}(\theta-\tilde{\theta}_j),
&\theta>\tilde{\theta}_j\end{array}\right. \label{eq:Lambda_rs2}
\end{align}
for $j=1,2$.

Hence, in order to comply with both the source and relay queueing constraints, the arrival rate $\aR_j$ at $\bS_j$ should satisfy (\ref{eq:cond1equ}) and (\ref{eq:cond2equ1}) simultaneously.

In this paper, system throughput is characterized by the pair of maximum constant arrival rates $\aR_1$ and $\aR_2$ that can be supported by the relay network with two pairs of source-destination nodes in the presence of statistical queueing constraints.

Finally, we provide a list of notations together with their descriptions in Table \ref{table1}.

\begin{table}
\begin{center}
\caption{\label{table1}Table of notations}
\begin{tabular}{|p{1cm}|p{7.5cm}|}
  \hline
  Notation & Definition\\
  \hline
  \hline
  $Y_j$ & Received signal at relay $\R$ (for $j=r$) or destination $\bD_j$ (for $j=1,2$). \\
  \hline
  $X_j$ & Transmitted signal from relay $\R$ (for $j=r$) or source $\bS_j$ (for $j=1,2$). \\
 \hline
  $g_j$ & Fading coefficient of the $\bS_j-\R$ link.\\
  \hline
  $z_j$ & Magnitude-square of the fading coefficient $g_j$.\\
  \hline
  $h_j$ & Fading coefficient of the $\R-\bD_j$ link.\\
  \hline
  $\omega_j$ & Magnitude-square of the fading coefficient $h_j$.\\
  \hline
  $n_j$ & Additive Gaussian noise at the relay $\R$( for $j=r$) or destination $\bD_j$ (for $j=1,2$) with variance $N_0$.\\
  \hline
  $\tsnr_j$ & Signal-to-noise ratio of relay $\R$ (for $j=r$) or source $\bS_j$ (for $j=1,2$).\\
  \hline
  $\theta_j$ & QoS exponent associated with the buffer constraint at relay $\R$ (for $j=r$) or source $\bS_j$ (for $j=1,2$).\\
  \hline
  $\Lambda(\theta)$ & LMGF of a departure or arrival process as a function of the QoS exponent $\theta$.\\
  \hline
  $\tau$ & The fraction of time allocated to the multiple-access phase.\\
  \hline
  $\rho$ & The fraction of power allocated by the relay to the transmission of the message intended for $\bD_1$.\\
  \hline
  $\delta$ & The fraction of time allocated to decoding order $\{1,2\}$ at relay $\R$.\\
  \hline
  $\aR_j$ & The maximum constant arrival rate at source $\bS_j$ that can be supported under queueing constraints.\\
  \hline
  $R_{A,B}$ & The instantaneous channel capacity of link $\mathbf{A}-\mathbf{B}$.\\
  \hline
  $r_{A,B}$ & The fixed transmission rate of link $\mathbf{A}-\mathbf{B}$ in the fixed rate scheme.\\
  \hline
\end{tabular}
\end{center}
\end{table}

\section{Throughput of the Two-Source Two-Destination Relay Network With Variable Transmission Rates}\label{varible_rate}
In this section, we study the throughput of the two-source two-destination relay network with variable-rate transmissions. Under the assumption that CSI is available at each transmitter, transmitters adapt their transmission rate to the instantaneous channel conditions, and the departure rates at each buffer are given by the corresponding instantaneous channel capacities. To perform an effective capacity analysis at each node with a buffer, we have to first identify the instantaneous transmission rates as functions of the fading coefficients.
\subsection{Instantaneous Transmission Rates in Multiple User Relay Networks}
We initially describe the instantaneous transmission rates of four links. Let us first consider the multiple-access phase in which links $\bS_1-\R$ and $\bS_2-\R$ are active simultaneously. When the decoding order at the relay is given by $\{1,2\}$, i.e., the information sent from node $\bS_1$ is decoded first, and the information sent from node $\bS_2$ is decoded after interference cancelation, then the maximum instantaneous achievable rates at $\bS_1$ and $\bS_2$ are given, respectively, by \cite{Deli_MAC}
\begin{align}
\begin{cases}\label{eq:MACrates12}
R_{\bS_1,\R\{1,2\}}&=B\log_2\left(1+\frac{\tsnr_1z_1}{1+\tsnr_2z_2}\right),\\
R_{\bS_2,\R\{1,2\}}&=B\log_2\left(1+\tsnr_2z_2\right).
\end{cases}
\end{align}
If the decoding order at the relay node is $\{2,1\}$, then we have
\begin{align}
\begin{cases}\label{eq:MACrates21}
R_{\bS_1,\R\{2,1\}}&=B\log_2\left(1+\tsnr_1z_1\right),\\
R_{\bS_2,\R\{2,1\}}&=B\log_2\left(1+ \frac{\tsnr_2z_2}{1+\tsnr_1z_1}\right).
\end{cases}
\end{align}
If we perform time-sharing between two decoding orders with parameter $\delta$, then the rates of links $\bS_1-\R$ and $\bS_2-\R$ are characterized by (\ref{eq:MACrates12}) in $\delta$ fraction of the time, and the rates are characterized by (\ref{eq:MACrates21}) rest of the time. Overall, the transmission rates between the source nodes and the relay node can be expressed as
\begin{equation}\label{eq:MACrates}
R_{\bS_j,\R}=\delta R_{\bS_j,\R\{1,2\}}+(1-\delta) R_{\bS_j,\R\{2,1\}},
\end{equation}
for $j=1,2$.

In the broadcast phase, relay node forwards packets to their corresponding destinations. In this phase, only links $\R-\bD_1$ and $\R-\bD_2$ are active. When the channel conditions are available at the relay node and destinations, the instantaneous transmission rates are given by
\begin{align}
\begin{cases}\label{eq:BCrates}
R_{\R,\bD_1}&=B\log_2\left(1+\frac{\rho\tsnr_r\omega_1}{1+(1-\rho)\tsnr_r\omega_1\mathds{1}\{\omega_1<\omega_2\}}\right), \\
R_{\R,\bD_2}&=B\log_2\left(1+\frac{(1-\rho)\tsnr_r\omega_2}{1+\rho\tsnr_r\omega_2\mathds{1}\{\omega_2<\omega_1\}}\right)
\end{cases}
\end{align}
where $\mathds{1}\{\bullet\}$ is indicator function.
\subsection{Stability Conditions}
With the expressions of the instantaneous rates for both the multiple-access channel and broadcast channel described above, we can characterize the stability region in the $\rho-\tau-\delta$ space. Stability at the source buffers is ensured by requiring the arrival rates to satisfy (\ref{eq:cond1equ}), which actually leads to compliance with the stricter condition that the tail distribution of the buffer length decays exponentially fast. The stability conditions at the relay node requires the average arrival rate to be less than or equal to the average departure rate at each buffer in the relay. Hence, the stability conditions can be formulated as
\begin{align}
\hspace{-0.5cm}
\begin{cases}\label{eq:stbcon1}
\tau\left(\delta\E\{R_{\bS_1,\R\{1,2\}}\}+(1-\delta)\E\{R_{\bS_1,\R\{2,1\}}\}\right)\\
\hspace{5.5cm}\leq(1-\tau)\E\{R_{\R,\bD_1}\},\\
\tau\left(\delta\E\{R_{\bS_2,\R\{1,2\}}\}+(1-\delta)\E\{R_{\bS_2,\R\{2,1\}}\}\right)\\
\hspace{5.5cm}\leq(1-\tau)\E\{R_{\R,\bD_2}\}.
\end{cases}
\end{align}
Plugging (\ref{eq:MACrates12}), (\ref{eq:MACrates21}), and (\ref{eq:BCrates}) into (\ref{eq:stbcon1}), we obtain
\begin{align}
\hspace{-0.5cm}
\begin{cases}\label{eq:stcon2}
(1-\tau)\E\left\{B\log_2\left(1+\frac{\rho\tsnr_r\omega_1}{1+(1-\rho)\tsnr_r\omega_1\mathds{1}\{\omega_1<\omega_2\}}\right)\right\}\geq\\
\hspace{2.7cm}\tau\bigg(\delta\E\{B\log_2\left(1+\frac{\tsnr_1z_1}{1+\tsnr_2z_2}\right)\}\\
\hspace{3.5cm}+(1-\delta)\E\{B\log_2\left(1+\tsnr_1z_1\right)\}\bigg),\\
(1-\tau)\E\left\{B\log_2\left(1+\frac{(1-\rho)\tsnr_r\omega_2}{1+\rho\tsnr_r\omega_2\mathds{1}\{\omega_2<\omega_1\}}\right)\right\}\geq\\
\hspace{2.7cm}\tau\bigg(\delta\E\{B\log_2\left(1+\tsnr_2z_2\right)\}\\
\hspace{3.2cm}+(1-\delta)\E\{B\log_2\left(1+\frac{\tsnr_2z_2}{1+\tsnr_1z_1}\right)\}\bigg).
\end{cases}
\end{align}
All feasible ($\rho$,$\tau$,$\delta$)-tuples satisfying the inequalities in (\ref{eq:stcon2}) form the stability region in the $\rho-\tau-\delta$ space. Hence, we formally define the the stability region $\Xi$ in the $\rho-\tau-\delta$ space as
\begin{equation}\label{stbregion1}
\Xi=\left\{(\rho,\tau,\delta)|\rho,\;\tau,\;\text{and}\;\delta\;\text{that satisfy (\ref{eq:stcon2})}\right\}.
\end{equation}
For a certain time sharing scheme at the relay node with fixed $\delta$, since $\tau$ is the time fraction allocated to the multiple-access phase, lower $\tau$ value is more likely to satisfy the stability condition, and the two inequalities in (\ref{eq:stcon2}) provide two upper bounds on $\tau$ as functions of $\rho$. The power allocation parameter $\rho$ has a different influence on these two phases. With more power allocated to transmission to $\bD_i$ in the broadcast phase, the corresponding buffer in the relay can support a higher $\tau$ value while satisfying the stability constraint.

\subsection{Throughput Region under Statistical Queueing Constraints}
As noted before, for a certain parameter setting, the system throughput is defined as the pair of constant arrival rates $\aR_1$ and $\aR_2$, which can be supported by two-hop links $\bS_1-\bD_1$ and $\bS_2-\bD_2$, respectively, under queueing constraints. Since stability is a prerequisite for effective capacity analysis, our system throughput is only defined with parameter values included in the stability region. For those parameter settings outside the stability region, at least one of the queueing constraints cannot be satisfied, and the system throughput is set to zero. Using the results in the previous section, to comply with queueing constraints at all nodes, $\aR_j$ for $j=1,2$ has to satisfy (\ref{eq:cond1equ}) and (\ref{eq:cond2equ1}) simultaneously, which leads to the following characterization of the system throughput.
\begin{Lem}
For any parameter setting $\{\tau,\rho,\delta\}$ that satisfies the stability conditions, the maximum constant arrival rate $\aR_j$, which can be supported at source node $\bS_j$ for $j = 1,2$ in the presence of all queueing constraints, is given by
\begin{align}
\hspace{-0.3cm}\aR_j=
\begin{cases}\label{throughput_1_2}
\min\Bigg\{-\frac{1}{\theta_j}\log(\E\{e^{-\theta_j\tau R_{\bS_j,\R}}\}),\\
\hspace{1.7cm}-\frac{1}{\theta_{r}}\log(\E\{e^{-\theta_r(1-\tau) R_{\R,\bD_j}}\})\Bigg\} \hspace{0.3cm}\theta_{r}\le \theta_j \\
\min\Bigg\{-\frac{1}{\theta_j} \log(\E\{e^{-\theta_j\tau R_{\bS_j,\R}}\}),\\
\hspace{1.5cm}-\frac{1}{\theta _j}\bigg(\log(\E\{e^{-\theta_r(1-\tau) R_{\R,\bD_j}}\})\\
\hspace{2.2cm}+\log(\E\{e^{(\theta_r-\theta_j)\tau R_{\bS_j,\R}}\})\bigg)\Bigg\}  \hspace{.3cm}\theta_{r}>\theta_j,
\end{cases}
\end{align}
\end{Lem}

\begin{IEEEproof}
We know that both $\bS_1-\bD_1$ and $\bS_2-\bD_2$ links are restricted by two queueing constraints, one at the corresponding source node, and the other one at the relay node. We consider these two constraints separately, and then combine the results. First, we only consider the constraints at the source nodes. According to (\ref{eq:cond1equ}), the maximum arrival rate that can be supported under queueing constraints at a source node is given by
\begin{align}\label{eq:rate_proof1}
\aR_j=-\frac{\Lambda_{\bS_j,\R}(-\theta_j)}{\theta_j},
\end{align}
for $j=1,2$.
Similarly, when we only consider the queueing constraint at the relay node, the maximum arrival rates should satisfy
\begin{align}
\hspace{-.3cm}\aR_j=
\begin{cases}\label{eq:rate_proof2}
-\frac{1}{\theta_{r}}\Lambda_{\R,\bD_j}(-\theta_r) &\theta_{r}\le \theta_j\\
-\frac{1}{\theta _j}\big(\Lambda_{\R,\bD_j}(-\theta_r)+\Lambda_{\bS_j,\R}(\theta_r-\theta_j)\big) &\theta_{r}>\theta_j,
\end{cases}
\end{align}
which is obtained from (\ref{eq:cond2equ1}) and (\ref{eq:Lambda_rs2}).
Combining these results, the overall maximum arrival rates that can be supported by the system should be the minimum of (\ref{eq:rate_proof1}) and (\ref{eq:rate_proof2}), i.e.,
\begin{align}
\hspace{-0.5cm}\aR_j=
\begin{cases}\label{throughput_1_1}
\min\Big\{-\frac{1}{\theta_j}\Lambda_{\bS_j,\R}(-\theta_j), -\frac{1}{\theta_{r}}\Lambda_{\R,\bD_j}(-\theta_r)\Big\} &\theta_{r}\le \theta_j \\
\min\Big\{-\frac{1}{\theta_j} \Lambda_{\bS_j,\R}(-\theta_j),\\
\hspace{0.8cm}-\frac{1}{\theta _j}\big(\Lambda_{\R,\bD_j}(-\theta_r)+\Lambda_{\bS_j,\R}(\theta_r-\theta_j)\big)\Big\} &\theta_{r}>\theta_j,
\end{cases}
\end{align}
for $j=1,2$.
Using the definition of LMGF in (\ref{LMGF_df}), (\ref{throughput_1_1}) can be expressed in terms of the instantaneous rates, which is given by (\ref{throughput_1_2}).

\end{IEEEproof}

Following this characterization, some properties of the system throughput are shown in the next subsection.

\subsection{Properties of the System Throughput under Queueing Constraints}
In the previous subsection, we have characterized the throughput of the two-source two-destination relay network. Based on (\ref{throughput_1_2}), we next analyze the behavior of the throughput in the parameter space, and establish several convexity properties, which can lead to simplifications in parameter optimization.

\begin{Lem}\label{the:2}
In the stability region, for a given $\tau-\rho$ pair, the maximum arrival rates $\aR_1$, $\aR_2$ and the sum rate $\aR_1+\aR_2$ are concave over the time sharing parameter $\delta$ between different decoding orders at the relay.
\end{Lem}

\begin{IEEEproof}
Depending on the relationship between $\theta_j$ and $\theta_r$ for $j=1,2$, there are two possible cases identified by (\ref{throughput_1_2}).\\
\\
$\mathbf{Case\quad 1}: \theta_{r}\le \theta_j$.\\
In this case, the throughput $\aR_j$ is given by
\begin{align}
\aR_j=\min\Bigg\{\aR_{j,1}, \aR_{j,2}\Bigg\}
\end{align}
where $\aR_{j,1}$ and $\aR_{j,2}$ are defined as
\begin{align}
\begin{cases}\label{eq:R_concave}
\aR_{j,1}=& -\frac{1}{\theta_j}\log\left(\E\left\{e^{-\theta_j\tau (\delta R_{\bS_j,\R\{1,2\}}+(1-\delta) R_{\bS_j,\R\{2,1\}})}\right\}\right),\\
\aR_{j,2}=& -\frac{1}{\theta_{r}}\log\left(\E\left\{e^{-\theta_r(1-\tau) R_{\R,\bD_j}}\right\}\right).
\end{cases}
\end{align}
By taking the second order derivative with respect to $\delta$, we can easily show the concavity of $\aR_{j,1}$ and $\aR_{j,2}$. The second order derivative of $\aR_{j,1}$ is given by (\ref{dR_j1}) on the next page.
\begin{figure*}
\small
\begin{align}
  \hspace{-.15cm}\frac{\partial^2 \aR_{j,1}}{\partial \delta^2}=-\frac{\theta_j\tau^2}{\bigg(\E\{e^{-\theta_j\tau (\delta R_{\bS_j,\R\{1,2\}}+(1-\delta) R_{\bS_j,\R\{2,1\}})}\}\bigg)^2}
  \Bigg\{\E\big\{(R_{\bS_j,\R\{1,2\}}-R_{\bS_j,\R\{2,1\}})^2 e^{-\theta_j\tau (\delta R_{\bS_j,\R\{1,2\}}+(1-\delta) R_{\bS_j,\R\{2,1\}})}\big\} \nonumber\\ \hspace{0.5cm}\E\{e^{-\theta_j\tau (\delta R_{\bS_j,\R\{1,2\}}+(1-\delta) R_{\bS_j,\R\{2,1\}})}\}-\bigg(\E\big\{(R_{\bS_j,\R\{1,2\}}-R_{\bS_j,\R\{2,1\}}) e^{-\theta_j\tau (\delta R_{\bS_j,\R\{1,2\}}+(1-\delta) R_{\bS_j,\R\{2,1\}})}\big\} \bigg)^2 \Bigg\}. \label{dR_j1}
\end{align}
\end{figure*}
\normalsize
According to the Cauchy-Schwarz inequality, two random variables $U$ and $V$ should satisfy $\E^2\{UV\}\leq\E\{U^2\}\E\{V^2\}$. Assuming that
\begin{equation}
U=e^{-\frac{1}{2}\theta_j\tau (\delta R_{\bS_j,\R\{1,2\}}+(1-\delta) R_{\bS_j,\R\{2,1\}})},
\end{equation}
and
\begin{equation}
V=(R_{\bS_j,\R\{1,2\}}-R_{\bS_j,\R\{2,1\}}) U,
\end{equation}
we can easily determine that the part inside the large curly brackets in (\ref{dR_j1}) can be written as $\E\{V^2\}E\{U^2\} - \E^2\{UV\}$ and hence is nonnegative. Then, we can readily determine that $\frac{\partial^2 \aR_{j,1}}{\partial \delta^2}\leq 0$, which indicates that $\aR_{j,1}$ is a concave function of $\delta$. From (\ref{eq:R_concave}), we notice that the expression of $\aR_{j,2}$ does not contain $\delta$. In other words, $\aR_{j,2}$ is a constant function in terms of $\delta$, and $\frac{\partial^2 \aR_{j,2}}{\partial \delta^2}= 0$. Hence, we can still regard $\aR_{j,2}$ as a concave function of $\delta$.

Since the pointwise minimum of concave functions is concave \cite{convex}, the concavity of $\aR_1$ and $\aR_2$ with respect to the time sharing parameter $\delta$ follows immediately when $\theta_{r}\le \theta_j$.\\
\\
$\mathbf{Case\quad 2}: \theta_{r}> \theta_j$.\\
In this case, the throughput $\aR_j$ is given by
\begin{align}
\aR_j=\min\Bigg\{\aR_{j,1}, \aR_{j,3}\Bigg\}
\end{align}
where $\aR_{j,3}$ is defined as

\begin{align}
\aR_{j,3}=-&\frac{1}{\theta _j}\bigg(\log(\E\{e^{-\theta_r(1-\tau) R_{\R,\bD_j}}\})\\ \notag
&+\log(\E\{e^{(\theta_r-\theta_j)\tau (\delta R_{\bS_j,\R\{1,2\}}+(1-\delta) R_{\bS_j,\R\{2,1\}})}\})\bigg).
\end{align}
We have already shown the concavity of $\aR_{j,1}$ in the previous case, and we can show the concavity of $\aR_{j,3}$ following the same approach. The second order derivative of $\aR_{j,3}$ is given by (\ref{dR_j3}) on the next page.
\begin{figure*}
\small
\begin{align}
\frac{\partial^2 \aR_{j,3}}{\partial \delta^2}=&-\frac{(\theta_r-\theta_j)^2\tau^2}{\theta_j \bigg(\E\{e^{(\theta_r-\theta_j)\tau (\delta R_{\bS_j,\R\{1,2\}}+(1-\delta) R_{\bS_j,\R\{2,1\}})}\}\bigg)^2} \nonumber
 \\
  &\hspace{-.9cm}\times\Bigg\{\E\big\{(R_{\bS_j,\R\{1,2\}}-R_{\bS_j,\R\{2,1\}})^2 e^{(\theta_r-\theta_j)\tau (\delta R_{\bS_j,\R\{1,2\}}+(1-\delta) R_{\bS_j,\R\{2,1\}})}\big\} \E\{e^{(\theta_r-\theta_j)\tau (\delta R_{\bS_j,\R\{1,2\}}+(1-\delta) R_{\bS_j,\R\{2,1\}})}\} \notag\\  &\hspace{-.3cm}-\bigg(\E\big\{(R_{\bS_j,\R\{1,2\}}-R_{\bS_j,\R\{2,1\}}) e^{(\theta_r-\theta_j)\tau (\delta R_{\bS_j,\R\{1,2\}}+(1-\delta) R_{\bS_j,\R\{2,1\}})}\big\} \bigg)^2 \Bigg\} \label{dR_j3}.
\end{align}
\end{figure*}
\normalsize
Again using the Cauchy-Schwarz inequality, we have $\frac{\partial^2 \aR_{j,3}}{\partial \delta^2}\leq 0$, and the concavity follows.
Since $\aR_j$ is the pointwise minimum of $\aR_{j,1}$ and $\aR_{j,3}$, $\aR_j$ is a concave function of $\delta$. Now, we have shown in both cases that $\aR_1$ and $\aR_2$ are concave functions of $\delta$.

Finally, since the sum of two concave functions is also a concave function, the sum rate is concave as well.
\end{IEEEproof}
Theorem \ref{the:2} indicates that there exists a globally optimal time sharing parameter for the two possible decoding orders at the relay, which can be determined via convex optimization methods. Similarly, the system throughput functions are also concave functions of $\tau$, which is the parameter for time allocation between the multiple-access and broadcast phases.

\begin{Lem}\label{the:3}
In the stability region, for given power allocation parameter $\rho$ and time-sharing parameter $\delta$, the maximum arrival rates $\aR_1$, $\aR_2$ and the sum rate $\aR_1+\aR_2$ are concave over the time allocation parameter $\tau$.
\end{Lem}

\begin{IEEEproof}
Similar to the proof of Theorem \ref{the:2}, Theorem \ref{the:3} can be proved easily by evaluating the derivatives with respect to $\tau$. The second order derivatives of $\aR_{j,1}$, $\aR_{j,2}$ and $\aR_{j,3}$ with respect to $\tau$ are given, respectively, by (\ref{dR_j1t})-(\ref{dR_j3t}) on the next page.
\begin{figure*}
\begin{align}
\label{dR_j1t}
\hspace{-0.0cm}\frac{\partial^2 \aR_{j,1}}{\partial \tau^2}=&-\frac{\theta_j}{\bigg(\E\{e^{-\theta_j\tau \R_{\bS_j,\R}}\}\bigg)^2} \bigg\{\E\{\R^2_{\bS_j,\R}e^{-\theta_j\tau \R_{\bS_j,\R}}\}\E\{e^{-\theta_j\tau \R_{\bS_j,\R}}\}-\bigg(\E\{\R_{\bS_j,\R}e^{-\theta_j\tau \R_{\bS_j,\R}}\}\bigg)^2\bigg\}
\end{align}
\begin{align}
\hspace{-0.0cm}\frac{\partial^2 \aR_{j,2}}{\partial \tau^2}=&-\frac{\theta_r}{\bigg(\E\{e^{-\theta_r(1-\tau) \R_{\R,\bD_j}}\}\bigg)^2} \notag
\\
&\hspace{.5cm}\times\bigg\{\E\{\R^2_{\R,\bD_j}e^{-\theta_r(1-\tau) \R_{\R,\bD_j}}\}\E\{e^{-\theta_r(1-\tau) \R_{\R,\bD_j}}\}-\bigg(\E\{\R_{\R,\bD_j}e^{-\theta_r(1-\tau) \R_{\R,\bD_j}}\}\bigg)^2\bigg\}  \label{dR_j2t}
\end{align}
\begin{align}
\hspace{-0.0cm}\frac{\partial^2 \aR_{j,3}}{\partial \tau^2}=&-\frac{\theta_r^2}{\bigg(\theta_j\E\{e^{-\theta_r(1-\tau) \R_{\R,\bD_j}}\}\bigg)^2} \nonumber
\\
&\hspace{.5cm}\times \bigg\{\E\{\R^2_{\R,\bD_j}e^{-\theta_r(1-\tau) \R_{\R,\bD_j}}\}\E\{e^{-\theta_r(1-\tau) \R_{\R,\bD_j}}\}-\bigg(\E\{\R_{\R,\bD_j}e^{-\theta_r(1-\tau) \R_{\R,\bD_j}}\}\bigg)^2\bigg\} \notag
\\
&-\frac{(\theta_r-\theta_j)^2}{\theta_j\bigg(\E\{e^{-\theta_j\tau \R_{\bS_j,\R}}\}\bigg)^2} \bigg\{\E\{\R^2_{\bS_j,\R}e^{-\theta_j\tau \R_{\bS_j,\R}}\}\E\{e^{-\theta_j\tau \R_{\bS_j,\R}}\}-\bigg(\E\{\R_{\bS_j,\R}e^{-\theta_j\tau \R_{\bS_j,\R}}\}\bigg)^2\bigg\}. \label{dR_j3t}
\end{align}
\end{figure*}
Using the Cauchy-Schwarz inequality and concavity-preserving property of pointwise minimum, the concavity of $\aR_1$, $\aR_2$ and the sum rate follow readily.
\end{IEEEproof}

Using these results, we can maximize the system throughput over $\delta$ and $\tau$ under stability constraints by employing efficient convex optimization methods.

\subsection{Numerical Results}
In this subsection, numerical results are provided to further analyze the throughput of the two-source two-destination relay network with variable transmission rates. Our numerical results are based on (\ref{throughput_1_2}).

\begin{figure}
\begin{center}
\centering
\includegraphics[width=0.45\textwidth]{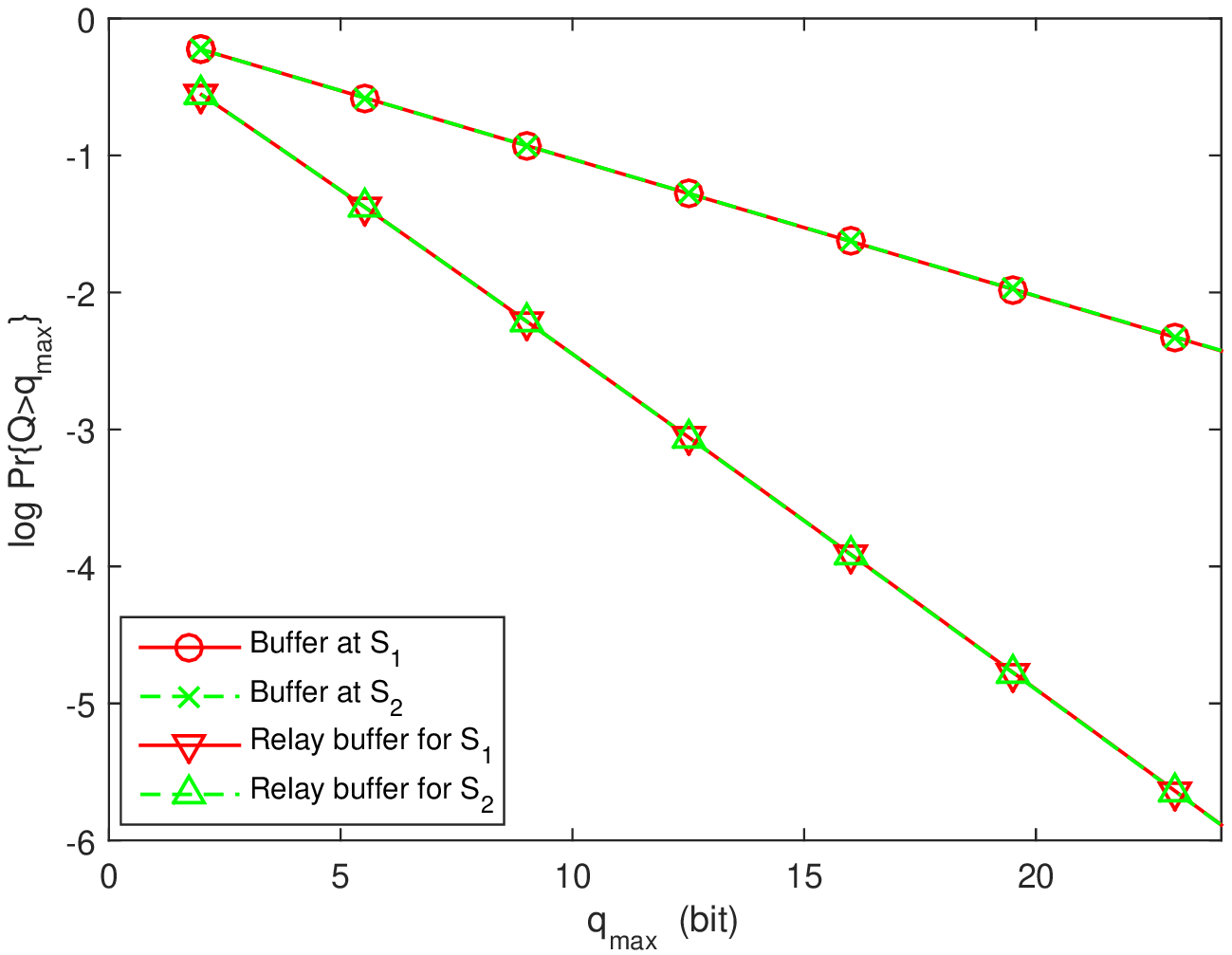}
\caption{Logarithmic buffer overflow probability vs. buffer overflow threshold.}\label{fig_sim1}
\end{center}
\end{figure}

\begin{figure}
\begin{center}
\includegraphics[width=0.45\textwidth]{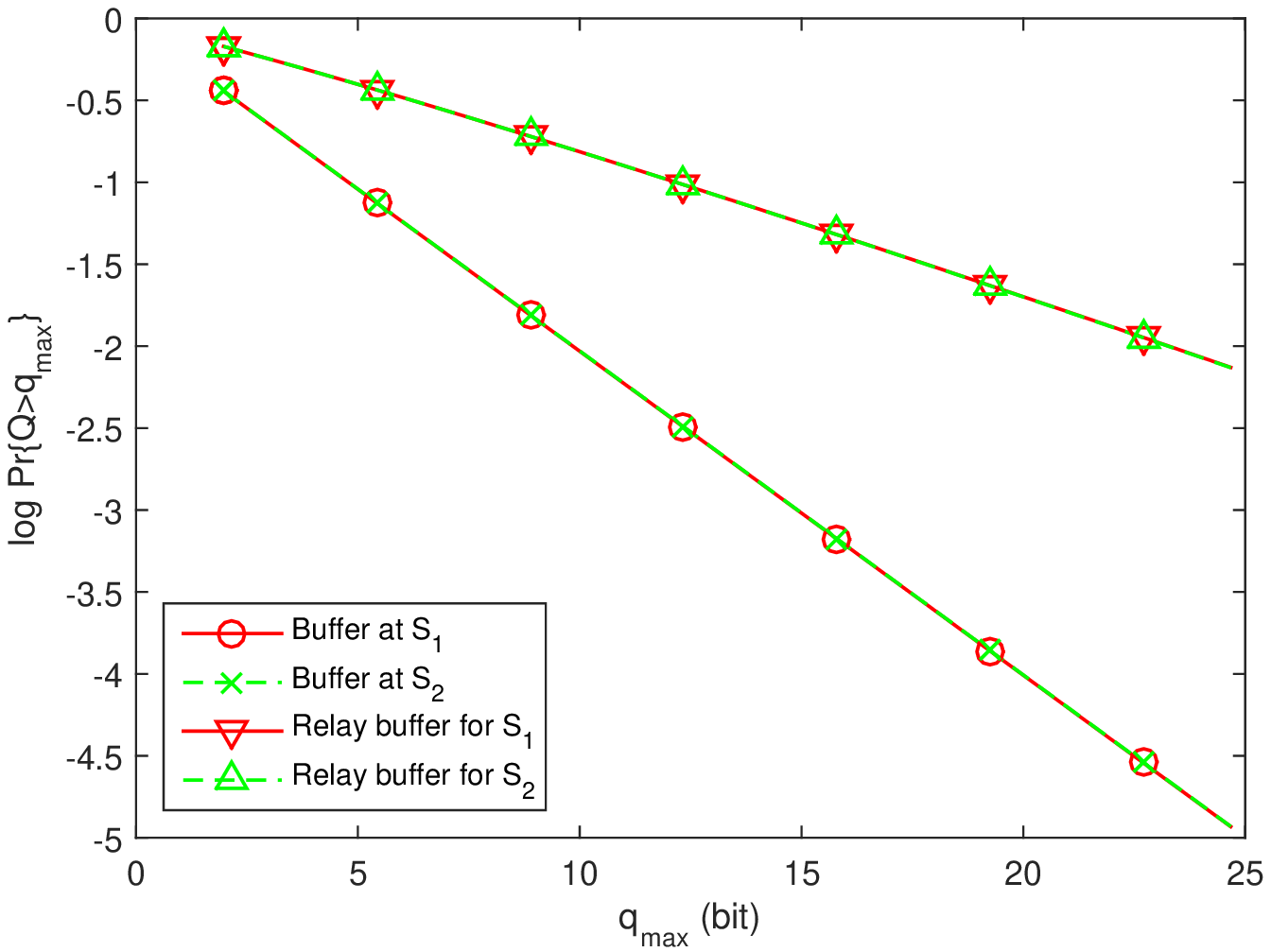}
\caption{Logarithmic buffer overflow probability vs. buffer overflow threshold.}\label{fig_sim2}
\end{center}
\end{figure}

In order to verify our analysis, we have conducted Monte Carlo simulations in which we have generated arrivals to the buffer at constant rates determined by our theoretical characterization in (\ref{throughput_1_2}) and also generated random (Rayleigh) fading coefficients to simulate the wireless channel and random transmission rates. We have tracked the buffer occupancy and overflows for different threshold levels.
We plot the simulated logarithmic buffer overflow probabilities as functions of the overflow threshold $q_{\max}$ in Figs. \ref{fig_sim1} and \ref{fig_sim2}. In each simulation, we generate $5\times 10^7$ time blocks to estimate the buffer overflow probability, and repeat each simulation $1000$ times to evaluate the averages. We set the queueing constraints as $\theta_1=\theta_2=\theta_r=0.1$, and the constant arrival rates at nodes $\bS_1$ and $\bS_2$ are determined from (\ref{throughput_1_2}). In both figures, $\E\{z_j\}=\E\{\omega_j\}=1$, $\tau=\rho=\delta=0.5$, $\tsnr_1=\tsnr_2=10$dB. In Fig. \ref{fig_sim1}, we set $\tsnr_r=30$dB. Note from (\ref{buffer_vio1}) that $\log \Pr\{Q \ge q_{\max}\} \approx \log \gamma -\theta q_{\max}$. Therefore, the slope of the logarithmic overflow probability is expected to be proportional to $-\theta$. Although (\ref{buffer_vio1}) requires large $q_{\max}$, our simulation results show that $\log \Pr\{Q \ge q_{\max}\}$ can be approximated as a linear function of $q_{\max}$ starting from relatively small $q_{\max}$. In Fig. \ref{fig_sim1}, the slopes of the logarithmic overflow probabilities at buffers in $\bS_1$ and $\bS_2$ are $-0.100$ and $-0.099$, respectively.
This implies that simulation results demonstrate perfect agreement with the analysis and the arrival rates given by (\ref{throughput_1_2}) fit the queueing constraints at $\bS_1$ and $\bS_2$ exactly. We also observe that the logarithmic overflow probabilities of the two relay buffers decay faster with steeper slopes than our requirement of $\theta_r = 0.1$. In this specific example, due to relay having a relatively large transmit power, the system performance is mainly decided by the multiple-access phase, which is the bottleneck of the system. Although the relay can potentially support higher $\aR_1$ and $\aR_2$, this is not allowed by the  multiple-access phase. As we reduce the transmission power of the relay node, the system bottleneck shifts to the broadcast phase and the situation is reversed. In Fig. \ref{fig_sim2}, we reduce $\tsnr_r$ to $27.5$dB. Now, the arrival rates given by (\ref{throughput_1_2}) fit the queueing constraints at the relay exactly, and the corresponding slopes are $-0.098$ and $-0.097$, respectively. On the other hand,  the decays of the overflow probabilities at the source nodes are faster, meaning that sources can potentially support higher arrival rates but this leads to the violation of the overflow constraints at the relay buffers and is therefore not allowed. Overall, these simulation results, while confirming the analysis, also interestingly unveil the critical interactions between the queues and buffer constraints.


\begin{figure}
\begin{center}
\includegraphics[width=0.45\textwidth]{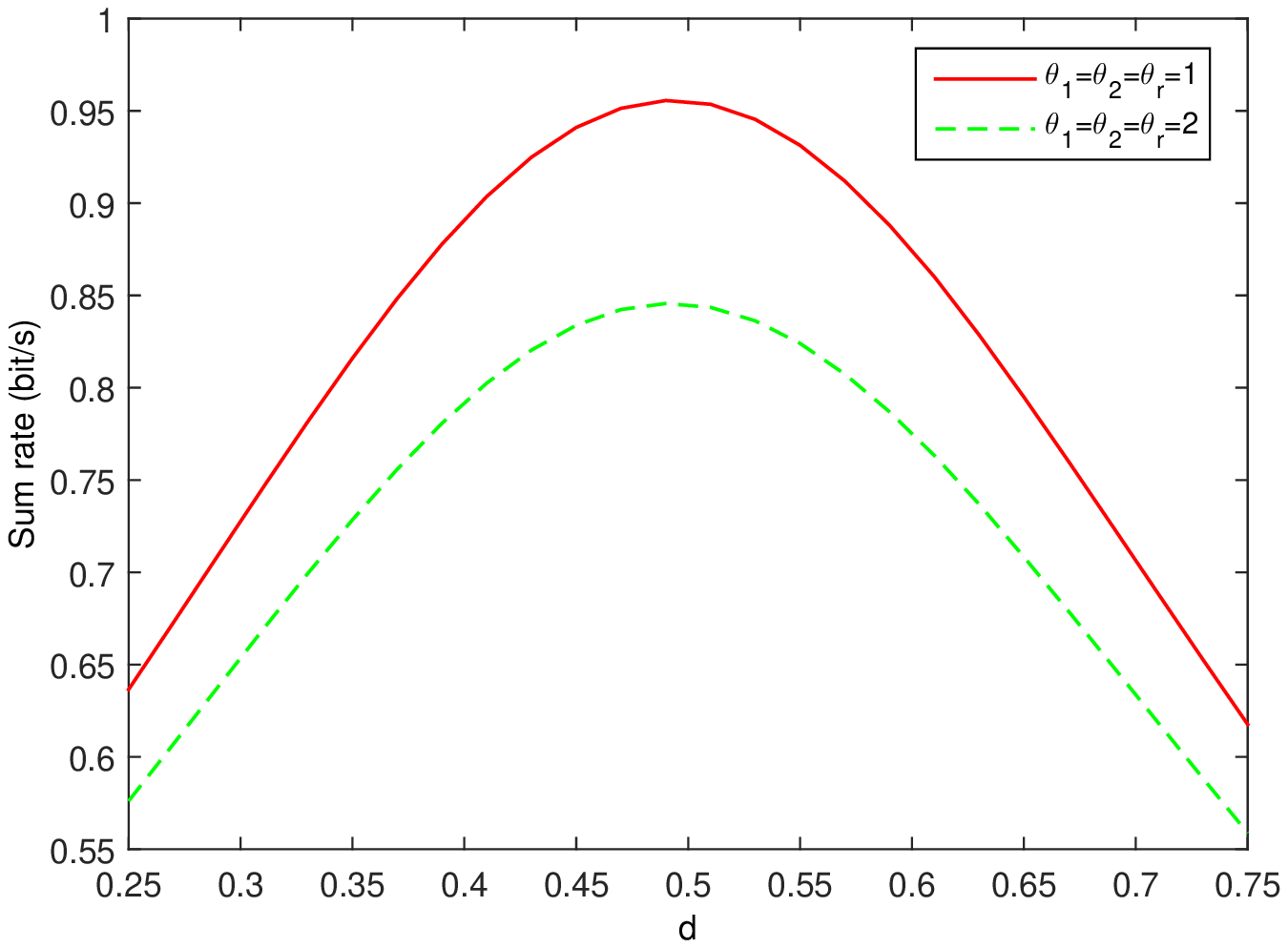}
\caption{The sum rate vs. relay location parameter.}\label{fig6}
\end{center}
\end{figure}

For the rest numerical results in this subsection, we consider Rayleigh fading and we set $\tsnr_1=\tsnr_2=3$dB and $\tsnr_r=6$dB. Fig. \ref{fig6} shows the influence of the position of the relay node for different $\theta$ values. We assume a symmetric model, in which $\theta_1 = \theta_2 = \theta_r$, and $Dist_{\bS_1,\R}=Dist_{\bS_2,\R}$ and $Dist_{\R,\bD_1}=Dist_{\R,\bD_2}$, where $Dist_{A,B}$ stands for the distance between $A$ and $B$. The overall distance $D=Dist_{\bS_1,\R}+Dist_{\R,\bD_1}=Dist_{\bS_2,\R}+Dist_{\R,\bD_2}=2$, and the position parameter $d=\frac{Dist_{\bS_1,\R}}{D}=\frac{Dist_{\bS_2,\R}}{D}$. Obviously, $d\in[0,1]$, and the smaller value of $d$ indicates that relay is closer to the source. Path loss as a function of distance is incorporated into the statistics of fading powers, and hence, we have $\E\{z_j\}=(\frac{1}{D\;d})^4$ and $\E\{\omega_j\}=(\frac{1}{D(1-d)})^4$ for $j=1,2$. In the figure, we see that the maximum sum rate $\aR_1+\aR_2$ is achieved when $d$ is close to $0.5$, which means that it is better to place the relay in the middle between the source and destination in this symmetric setting. When the relay is close to the source nodes, the channels between the relay and destinations deteriorate and the overall throughput is limited by the broadcast links. Similarly, the multiple-access links become the bottleneck when $d$ is close to $1$. Also, we observe that the system throughput decreases when $\theta$ increases due to tighter queueing constraints. This occurs because when $\theta$ is small, the effective capacity is closer to the Shannon capacity, and as $\theta$ increases, effective capacity diminishes and approaches the zero-outage capacity (which is, for instance, zero in Rayleigh fading).

\begin{figure}
\begin{center}
\includegraphics[width=0.45\textwidth]{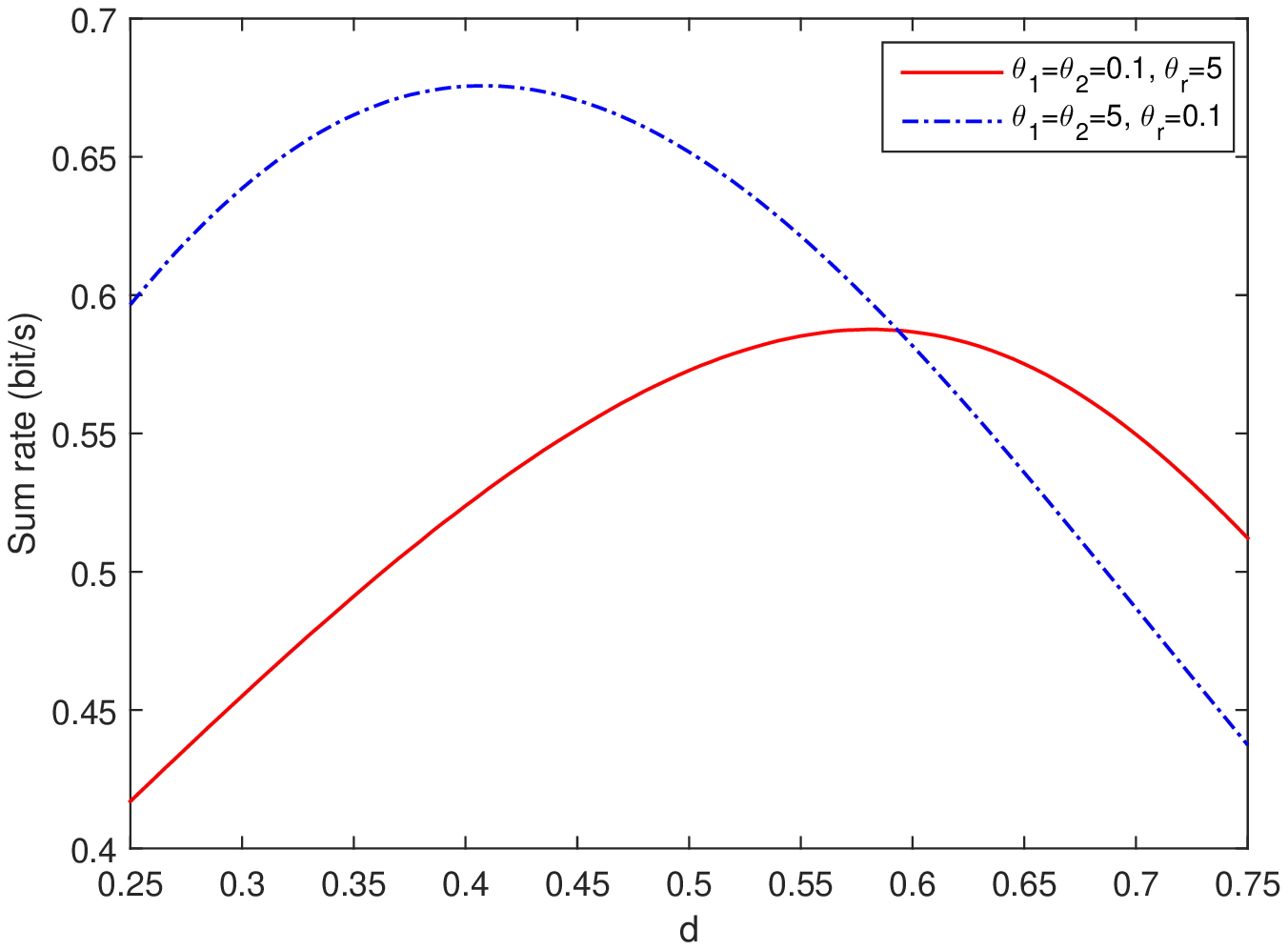}
\caption{The sum rate vs. relay location parameter.}\label{fig_NP}
\end{center}
\end{figure}

In Fig. \ref{fig_NP}, we consider an asymmetric scenario in terms of QoS exponents, and again plot sum rate vs. relay location parameter $d$.  We fix $\rho=\delta=0.5$ and determine the optimal value of $\tau$ for each given $d$. When $\theta_r=5$, $\theta_1=\theta_2=0.1$, the maximum sum rate is achieved at $d=0.58$. In this case, relay should be placed closer to the destinations to support more stringent queueing constraints at the relay. On the other hand, when $\theta_1=\theta_2=5$, $\theta_r=0.1$, the optimal position for the relay is at $d=0.41$. Hence, the relay needs to be closer to the source nodes to support their stricter queueing constraints. These observations indicate the sensitivity of optimal relay placement to different QoS requirements.

\begin{figure}
\begin{center}
\centering \includegraphics[width=0.45\textwidth]{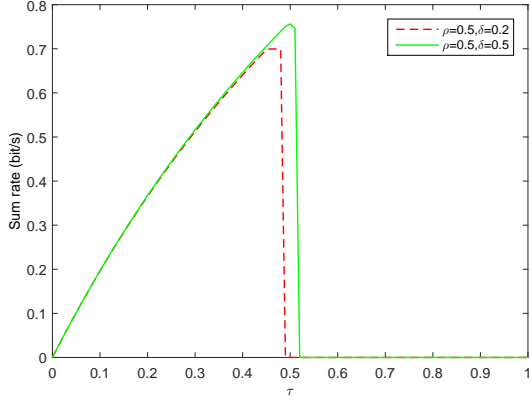}
\caption{The sum rate vs. time allocation parameter $\tau$.}\label{fig7}
\end{center}
\end{figure}

\begin{figure}
\begin{center}
\centering \includegraphics[width=0.45\textwidth]{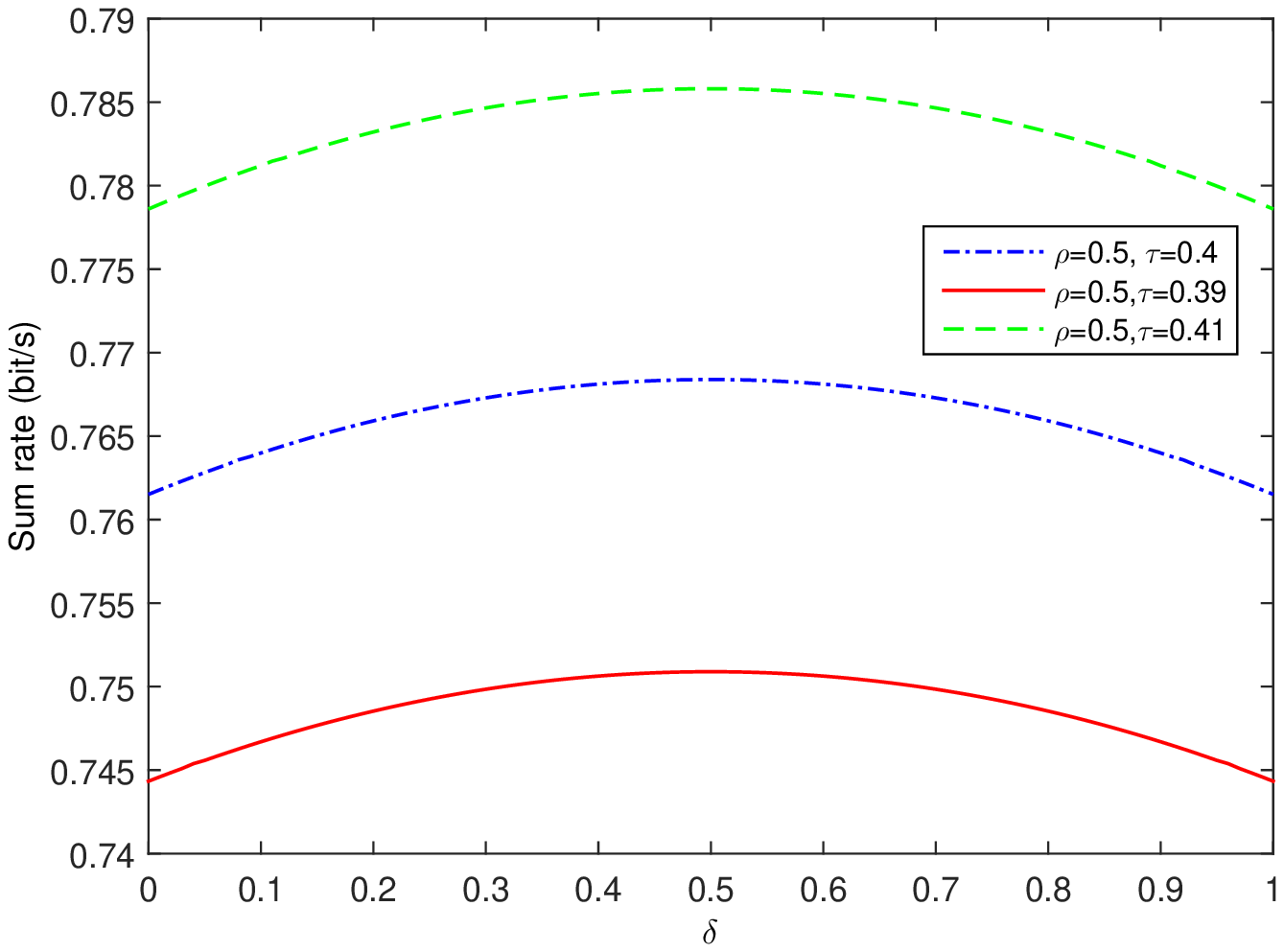}
\caption{The sum rate vs. decoding parameter $\delta$.}\label{fig8}
\end{center}
\end{figure}

\begin{figure}
\begin{center}
\includegraphics[width=0.45\textwidth]{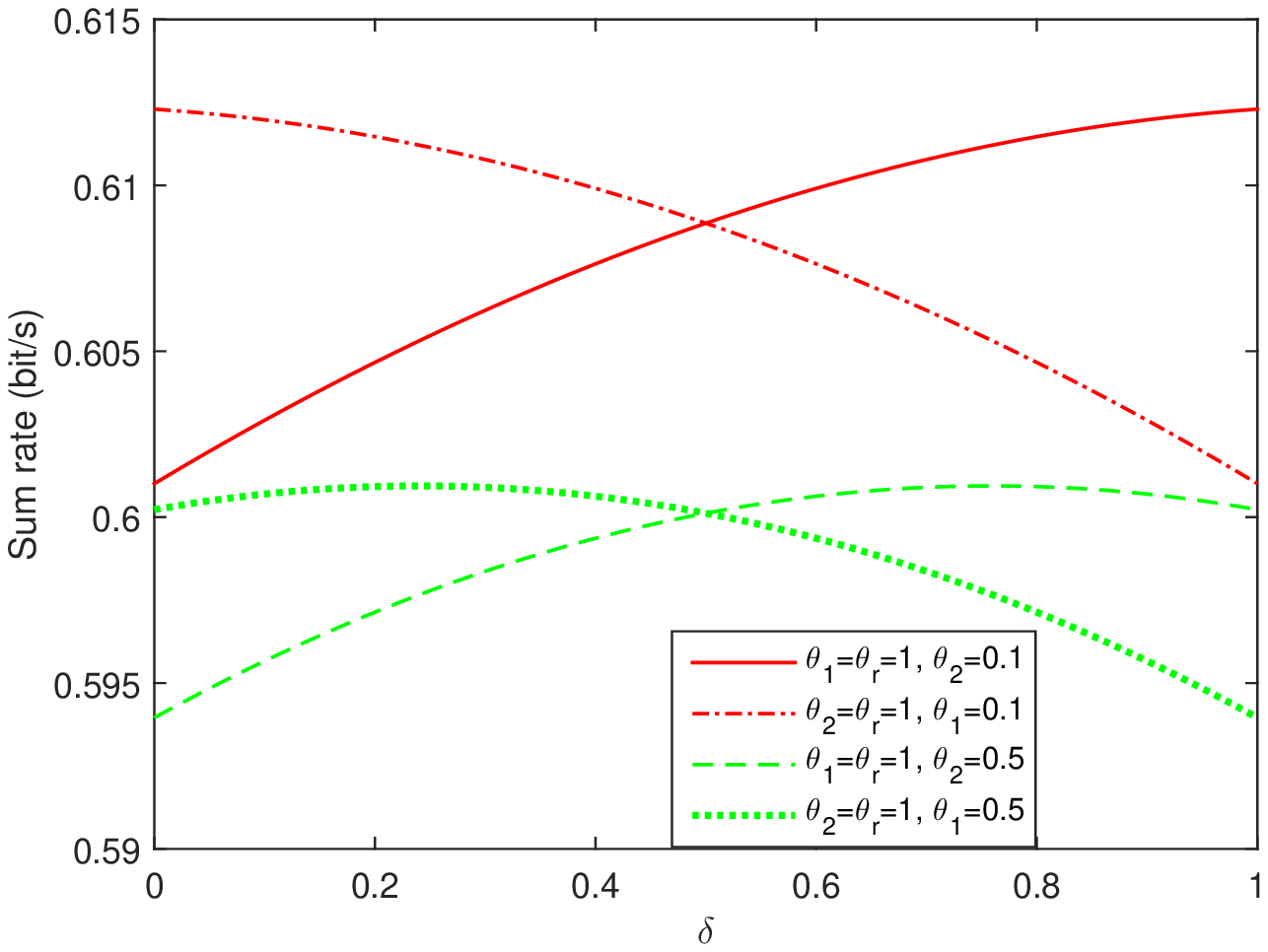}
\caption{The sum rate vs. decoding parameter $\delta$.}\label{fig-sum-delta-new}
\end{center}
\end{figure}

Figs. \ref{fig7} and \ref{fig8} demonstrate the concavity\footnote{These concavity results can simplify the search for the optimal parameter setting with the use of convex optimization tools.} of the sum rate with respect to $\tau$ and $\delta$, respectively, when the parameter values are in the stability region. In these two figures, $\theta_1=\theta_2=\theta_r=1$, and $\E\{z_j\}=\E\{\omega_j\}=1$. In Fig. \ref{fig7}, the sum rate curves first increase with $\tau$, and then decrease very fast after reaching the maximum sum rate. As $\tau$ exceeds a threshold, the sum rates drop to $0$, because stability conditions are violated beyond this threshold. In Fig. \ref{fig8}, the sum rate curves are concave with respect to the decoding parameter $\delta$, and the optimal $\delta$ values which maximize the sum rate are all close to $0.5$. In this case, relay allocates time to two decoding orders equally. However, note that these results are again for a symmetric scenario in which all QoS exponents are the same. In Fig. \ref{fig-sum-delta-new}, we address a heterogeneous setting in terms of QoS exponents. For instance, when $\theta_1=\theta_r=1$ and $\theta_2 = 0.1$, the optimal value of $\delta$ is 1. Hence, sum rate is maximized when the decoding order at the relay is always fixed as $\{1,2\}$, i.e., relay initially decodes data arriving from source $\bS_1$ in the presence of interfering signal of $\bS_2$. The underlying reason for this result is the following. Source $\bS_1$ operates under stricter QoS constraints with respect to $\bS_2$ and consequently can support smaller arrival rates and needs, in turn, smaller transmission rates which can be sustained even in the presence of interference. If the roles are switched (i.e., if we have $\theta_2=\theta_r=1$ and $\theta_1 = 0.1$), then the optimal value of $\delta$ is zero. If the QoS exponents are more comparable (e.g., $\theta_1 = 1$ and $\theta_2 = 0.5$ or $\theta_1 = 0.5$ and $\theta_2 = 1$), we notice that optimal values of $\delta$ start to slightly deviate from the two extremes of 0 and 1.

\begin{figure}
\begin{center}
\includegraphics[width=0.45\textwidth]{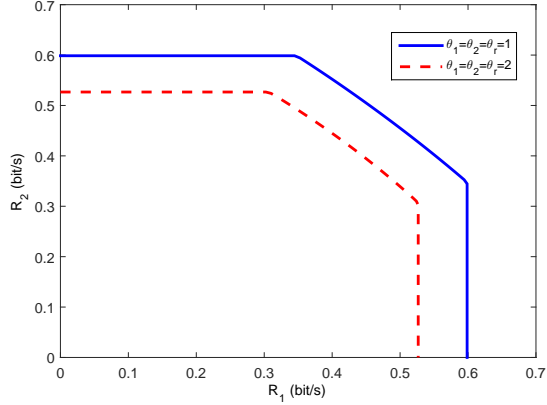}
\caption{System throughput region $\aR_1$ vs. $\aR_2$.}\label{fig9}
\end{center}
\end{figure}

Fig. \ref{fig9} shows the throughput regions of the two-source two-destination relay network under different queueing constraints. The boundary of the throughput region is obtained by searching over the three-dimensional parameter space. When $\aR_1$ achieves its maximum value, $\delta$ is close to $0$, and $\rho$ is slightly greater than $0.5$, because decoding order $\{2,1\}$ and more power in the $\R-\bD_1$ link can help $\bS_1-\bD_1$ link to support higher arrival rates. Similar results are also obtained for the maximum value of the arrival rate $\aR_2$.

\section{Extension to Multiple Source-Destination and Full-Duplex Models}
In this section, we study the extensions of the two-source two-destination relay model, addressed in Section \ref{varible_rate}. The first extension is to generalize the results to the multi-source multi-destination relay model with more than two source-destination pairs. The second extension is to full-duplex operation of the relay.

\subsection{Multi-Source Multi-Destination Relay Network}
In this subsection, we consider a multiple-user model in which $N$ sources send information to their corresponding destinations with the help of a relay node. The magnitude-squares of the fading coefficients of links $\bS_j-\R$ and $\R-\bD_j$ are represented by $z_j$ and $\omega_j$, respectively. We assume that there are $N$ separate buffers in the relay node, each one for the information arriving from a different source. Again, all buffers in the relay node are assumed to share the same QoS exponent $\theta_r$, while source node $\bS_j$ may have its own QoS exponent $\theta_j$. Compared with the two-user model, adding more users only increases the dimension of the parameter space while the analytical methods and results essentially remain the same.

In this multi-user setting, system parameters $\rho$ and $\delta$ become vectors, while the time allocation parameter $\tau$ is still a scalar. The definition of $\tau$ is kept the same as the fraction of time allocated to the multiple-access phase. The power allocation parameter $\rho$ becomes an $N$-dimensional vector $\boldsymbol{\rho}=(\rho_1,\rho_2,\cdots,\rho_N)$, and the $i^{\text{th}}$ component $\rho_i$ represents the fraction of power allocated to the data transmission to $\bD_i$, $i=1,2,\cdots N$. The elements of $\boldsymbol{\rho}$ should be between $0$ and $1$, and satisfy $\sum_{i=1}^{N}\rho_i=1$. Since there are $N!$ different decoding orders at the relay in the multiple-access phase, the time-sharing parameter $\delta$ becomes an $N!$-dimensional vector $\boldsymbol{\delta}=(\delta_1,\delta_2,\cdots,\delta_{N!})$, and the $i^{\text{th}}$ component $\delta_i$ represents the fraction of time allocated to the $i^{\text{th}}$ decoding order at the relay node. Similarly, all the elements of $\boldsymbol{\delta}$ should be between $0$ and $1$, and satisfy $\sum_{i=1}^{N!}\delta_i=1$.

In the multiple-access phase, we denote the $k^{\text{th}}$ decoding order at the relay as $\Or_k=\{k_1,k_2,\cdots,k_N\}$, which is a permutation of $\{1,2,\cdots,N\}$. With this decoding order, the instantaneous rate of the $\bS_{k_i}-\R$ link is characterized by

\begin{equation}
\R_{\bS_{k_i},\R,\Or_k}=B\log_2\left(1+ \frac{\tsnr_{k_i}z_{k_i}}{1+\sum_{j=i+1}^{N}\tsnr_{k_j}z_{k_j}}\right).
\end{equation}
Given a time sharing vector $\boldsymbol{\delta}=(\delta_1,\delta_2,\cdots,\delta_{N!})$, the rate of the $\bS_j-\R$ link is given by

\begin{equation}\label{MCrate_multi-user}
\R_{\bS_j,\R}=\sum_{k=1}^{N!}\delta_k \R_{\bS_j,\R,\Or_k},
\end{equation}
for $j=1,2,\cdots,N$. For the broadcast channel, the instantaneous rate is given by
\begin{equation}\label{BCrate_multi-user} R_{\R,\bD_j}=B\log_2\left(1+\frac{\rho_j\tsnr_r\omega_j}{1+\sum_{l=1,l\neq j}^{N}\rho_l\tsnr_r\omega_j\mathds{1}\{\omega_j<\omega_l\}}\right),
\end{equation}
for $j=1,2,\cdots,N$.
Similarly, the stability region in the parameter space is defined as

\begin{align}\label{stability_2}
\Xi=\bigg\{(\tau,&\rho_1,\cdots,\rho_N,\delta_1,\cdots,\delta_{N!})|\tau,\boldsymbol{\rho}\;\text{and}\; \boldsymbol{\delta}\;
            \text{that satisfy}\notag\\
&\hspace{1.5cm}\tau\E\{\R_{\bS_j,\R}\}\leq(1-\tau)\E\{R_{\R,\bD_j}\},\notag\\
         &\sum_{i=1}^{N}\rho_i=1\;\text{and}\;\sum_{i=1}^{N!}\delta_i=1, \text{for all}\;j=1,2,\cdots,N\bigg\}.
\end{align}
In this multiple-user setting, the dimension of the parameter space becomes much higher than that in the two-user model.
For a set of parameters that guarantee the stability conditions, the throughput of the $\bS_j-\bD_j$ link under queueing constraints satisfies (\ref{eq:cond1equ}) and (\ref{eq:cond2equ1}) simultaneously, and hence is given by (\ref{throughput_1_2}), for $j=1,2,\cdots,N$, with the instantaneous rate expressions provided above.

\subsection{Full-Duplex Two-Source Two-Destination Relay Network}
In this subsection, we extend our analysis on half-duplex system to full-duplex two-source two-destination relay network. In full-duplex mode, nodes $\bS_1$, $\bS_2$ and relay transmit all the time, and thus there is no parameter $\tau$ in this case. In full-duplex mode, relay node experiences self-interference due to its transmission to $\bD_1$ and $\bD_2$. The received signal at relay is given by
\begin{equation}
Y_r[i]=g_1[i]X_1[i]+g_2[i]X_2[i]+I_s[i]+n_r[i]
\end{equation}
where $I_s[i]$ is the self-interference term caused by simultaneous transmission of the signal that relay sends to $\bD_1$ and $\bD_2$. Since relay knows the signal it sends to the destination nodes, it can perform self-interference cancelation. Here, we assume that relay can perfectly eliminate its self-interference, and the received signal at relay after self-interference cancelation is given by (\ref{eq:RSG_R}). In order to satisfy the stability conditions at the relay, two source nodes need to reduce their transmission power, when the arrival rates are larger than departure rates. Therefore, we introduce two new parameters, $\alpha_1$ and $\alpha_2$, which represent the fraction of power that $\bS_1$ and $\bS_2$ use for transmission, respectively. Obviously, we have $\alpha_1, \alpha_2 \in [0,1]$. Parameters $\rho$ and $\delta$ are defined in the same way as in the half-duplex mode. Then, for a given $(\alpha_1,\alpha_2)$ pair, the instantaneous transmission rates in the multiple-access phase become
\begin{align}
\begin{cases}\label{eq:FD_MACrates12}
R_{\bS_1,\R\{1,2\}}&=B\log_2\left(1+\frac{\alpha_1\tsnr_1z_1}{1+\alpha_2\tsnr_2z_2}\right),\\
R_{\bS_2,\R\{1,2\}}&=B\log_2\left(1+\alpha_2\tsnr_2z_2\right),
\end{cases}
\end{align}
if the decoding order at the relay node is $\{1,2\}$. On the other hand, we have
\begin{align}
\begin{cases}\label{eq:FD_MACrates21}
R_{\bS_1,\R\{2,1\}}&=B\log_2\left(1+\alpha_1\tsnr_1z_1\right),\\
R_{\bS_2,\R\{2,1\}}&=B\log_2\left(1+ \frac{\alpha_2\tsnr_2z_2}{1+\alpha_1\tsnr_1z_1}\right),
\end{cases}
\end{align}
when the decoding order is $\{2,1\}$. Plugging (\ref{eq:FD_MACrates12}) and (\ref{eq:FD_MACrates21}) into (\ref{eq:MACrates}), we can get the overall transmission rates between sources and relay, and the transmission rates in the broadcast phase are also given by (\ref{eq:BCrates}).

In order to satisfy the stability conditions at the relay node, we have to choose a parameter setting that can satisfy

\begin{align}
\begin{cases}\label{eq:stbcon3}
  \E\{R_{\bS_1,\R}(\alpha_1,\alpha_2,\delta)\}& \leq \E\{R_{\R,\bD_1}(\rho)\}\\
  \E\{R_{\bS_2,\R}(\alpha_1,\alpha_2,\delta)\}& \leq \E\{R_{\R,\bD_2}(\rho)\},
\end{cases}
\end{align}
and the stability region in full-duplex mode becomes
\begin{align}
  \Xi=\left\{(\alpha_1,\alpha_2,\tau,\delta)|\alpha_1,\;\alpha_2\;\tau,\;\text{and}\;\delta\;\text{that satisfy (\ref{eq:stbcon3})}\right\}.
\end{align}

Through a similar process as in \cite[Theorem 2]{deli-relay}, the constant arrival rates at source nodes, $\aR_j$ for $j=1,2$, are given by the following:
\begin{enumerate}
  \item If $\theta_j\geq\theta_r$,

    \begin{align}
      \aR_j=&\min\bigg\{-\frac{1}{\theta_j}\log\E\{e^{-\theta_j R_{\bS_j,\R}}\},\notag\\
 &\hspace{2.5cm}-\frac{1}{\theta_r}\log\E\{e^{-\theta_r R_{\R,\bD_j}}\}\bigg\}.
    \end{align}

  \item If $\theta_j\leq\theta_r$ and $\theta_r\leq\bar{\theta}$,

\begin{align}
 \aR_j=-\frac{1}{\theta_j}\log\E\{e^{-\theta_j R_{\bS_j,\R}}\}
\end{align}
where $\bar{\theta}$ is the $\theta$ value that makes the following equation satisfied:
\begin{align}
\hspace{-0.5cm} -\frac{1}{\theta_j}\log\E\{e^{-\theta_j R_{\bS_j,\R}}\}=-\frac{1}{\theta_j}\bigg(&\log\E\{e^{-\theta R_{\R,\bD_j}}\}\notag\\
&+\log\E\{e^{(\theta-\theta_j)R_{\bS_j,\R}}\}\bigg).
\end{align}
  \item $\theta_j\leq\theta_r$ and $\theta_r\geq\bar{\theta}$
\begin{enumerate}
  \item If

    \begin{align}
         -\frac{1}{\theta_r}\log\E\{e^{-\theta_r R_{\R,\bD_j}}\}\geq -\frac{1}{\theta_r}\log\E\{e^{-\theta_r R_{\bS_j,\R}}\},
    \end{align}
    then

    \begin{align}
      \aR_j=-\frac{1}{\theta^*}\log\E\{e^{-\theta^* R_{\bS_j,\R}}\}
    \end{align}
    where $\theta^*$ is the smallest solution to
    \begin{align}
      \hspace{-0.5cm}-\frac{1}{\theta}\log\E\{e^{-\theta R_{\bS_j,\R}}\}=-\frac{1}{\theta}\bigg(&\log\E\{e^{-\theta_r R_{\R,\bD_j}}\}\notag\\
&+\log\E\{e^{(\theta_r-\theta)R_{\bS_j,\R}}\}\bigg).
    \end{align}

  \item If

    \begin{align}
        -\frac{1}{\theta_r}\log\E\{e^{-\theta_r R_{\R,\bD_j}}\}<-\frac{1}{\theta_r}\log\E\{e^{-\theta_r R_{\bS_j,\R}}\}
    \end{align}
    and

    \begin{align}
      -\frac{1}{\theta_r}\log\E\{e^{-\theta_r R_{\R,\bD_j}}\}\geq \inf_{z_1,z_2} R_{\bS_j,\R}
    \end{align}
    then

    \begin{align}
      \aR_j=-\frac{1}{\theta^*}\log\E\{e^{-\theta^* R_{\bS_j,\R}}\}
    \end{align}
    where $\theta^*$ is the solution to
    \begin{align}
      -\frac{1}{\theta}\log\E\{e^{-\theta R_{\bS_j,\R}}\}=-\frac{1}{\theta_r}\log\E\{e^{-\theta_r R_{\R,\bD_j}}\}.
    \end{align}
  \item Otherwise

    \begin{align}
      \aR_j=-\frac{1}{\theta_r}\log\E\{e^{-\theta_r R_{\R,\bD_j}}\}.
    \end{align}

\end{enumerate}
\end{enumerate}

\section{Throughput of the Two-Source Two-Destination Relay Network With Fixed Transmission Rates}
In practice, CSI may not be available at the transmitters. In such cases, the instantaneous departure rates from each buffer will be different. In this section, we investigate the system throughput when the transmitters do not have CSI and transmit at fixed rates. We further assume that an ARQ protocol is employed and retransmissions are requested in case of communication failures. \cite{ARQ_fix}

In the ARQ protocol, if the receiver decodes the packet, an ACK feedback is sent to the transmitter, otherwise the receiver asks for the retransmission of the same packet until the receiver gets the packet correctly. Here, the feedback signals are assumed to be transmitted without error and delay. In other words, the transmitter gets the error free feedback signal immediately after it completes the transmission of the corresponding packet. In this model, ARQ scheme guarantees the reliability, and the packets are kept in the buffer until the receiver decodes it correctly. With this ARQ assumption, the instantaneous departure rate at a buffer is equal to the fixed transmission rate if the receiver decodes the packet correctly, and it is $0$ if the transmission fails.

In order to determine the asymptotic LMGFs $\Lambda_{\bS_j, \R}$, $\Lambda_{\R}$, and $\Lambda_{\R,\bD_j}$, we have to first identify the success and failure probabilities of these fixed-rate transmissions.

\subsection{State Probabilities in the Multiple Access Phase}

As noted before, source node $\bS_j$ transmits in the multiple-access phase with fixed rate $r_{\bS_j,\R}$ for $j=1,2$. In the broadcast phase, relay node transmits to destination $\bD_j$ with fixed rate $r_{\R,\bD_j}$, for $j=1,2$. Since all transmitters are using the ARQ protocol, all links can be regarded to be in either ON or OFF state at a given time. The link is in the ON state if the fixed transmission rate is less than the instantaneous channel capacity, and the receiver can decode the packet correctly. Otherwise, failure occurs and the link is in the OFF state in which the transmission rate is effectively zero.

In the multiple-access phase, the channel capacity is related to the decoding strategy of the relay, which is described as follows:
\begin{enumerate}
  \item Relay tries to decode the first packet while treating the interference as noise. Without loss of generality, we assume that the relay always starts with the packets sent by $\bS_1$.
  \begin{enumerate}
    \item If the receiver decodes the packet correctly, then it moves to the interference cancelation step (i.e., Step 2 below).
    \item If the receiver cannot decode the packet from $\bS_1$, it tries to decode the packet from $\bS_2$.
    \item If the receiver decodes it correctly, then it moves to the interference cancelation step. Otherwise, it asks retransmission from both transmitters, and decoding process ends.
  \end{enumerate}
  \item The receiver performs interference cancelation by subtracting the decoded message from the received signal.
  \item The receiver attempts to decode the remaining packet after interference cancelation. If it cannot decode the packet, retransmission is required from the corresponding transmitter.
\end{enumerate}
Later in our analysis, we show that it does not make any difference if the relay starts with the packets sent by $\bS_2$. In the multiple-access phase, according to the states of the links $\bS_1-\R$ and $\bS_2-\R$, we identify four possible cases:
\\
\\
\textbf{Case 1}: In this case, the relay node cannot decode any of the received messages. Relay node attempts to decode the message from $\bS_1$ first, while treating the signal from $\bS_2$ as noise. Following unsuccessful decoding, relay tries to decode the message from $\bS_2$ while treating the interference as noise, and cannot succeed either. Hence, we in this scenario have

\begin{align}
\begin{cases}\label{case1_1}
r_{\bS_1,\R}>&\tau B\log_2\left(1+\frac{\tsnr_1 z_1}{1+\tsnr_2 z_2}\right)\\
r_{\bS_2,\R}>&\tau B\log_2\left(1+\frac{\tsnr_2 z_2}{1+\tsnr_1 z_1}\right)
\end{cases}.
\end{align}
(\ref{case1_1}) can be transformed into the following bounds on fading magnitude-squares $z_1$ and $z_2$:
\begin{small}
\begin{align}
\begin{cases}\label{case1_2}
z_1>&\frac{1}{\tsnr_1}\left(\tsnr_2 z_2/\left(\prb-1\right)-1\right)\\
z_1<&\frac{1}{\tsnr_1}\left(\pra-1\right)(1+\tsnr_2 z_2)\\
z_2>&0\\
z_2<&-\left(\prb-1\right)\pra\bigg/\\
    &\qquad\left\{\tsnr_2\left[\left(\pra-1\right)\left(\prb-1\right)-1\right]\right\},\\
    &\qquad\qquad\quad\text{if} \left(\pra-1\right)\left(\prb-1\right)<1.
\end{cases}
\end{align}
\end{small}
\hspace{-.3cm} (\ref{case1_2}) defines a region on the first quadrant of $(z_1,z_2)$ plane, which we denote by $\Psi_1$. Therefore, the probability of Case $1$ is given by

\begin{equation}\label{pm1}
P_{M,1}=\iint\limits_{\Psi_1}p_{z_1,z_2}(z_1,z_2)dz_1dz_2,
\end{equation}
where $p_{z_1,z_2}(z_1,z_2)$ is the joint probability density function (pdf) of $z_1$ and $z_2$. For instance, if we consider independent Rayleigh fading, then joint pdf is given by
\begin{equation}\label{zpdf}
 p_{z_1,z_2}(z_1,z_2)=\frac{1}{\overline{z_1}\:\overline{z_2}}\exp\left(-\frac{z_1}{\overline{z_1}}-\frac{z_2}{\overline{z_2}}\right),
\end{equation}
where $\overline{z_j}$ represents the expected value of $z_j$ for $j=1,2$.

In this case, since the relay can decode none of them, switching the decoding order will not make a difference.
\\
\\
\textbf{Case 2}: In this case, the relay can decode the message from $\bS_1$ in the presence of interference from $\bS_2$, but the message from $\bS_2$ cannot be decoded successfully even after interference cancelation. This scenario can be expressed by the following two inequalities:
\begin{align}
\begin{cases}\label{case2_1}
r_{\bS_1,\R}\leq&\tau B\log_2\left(1+\frac{\tsnr_1 z_1}{1+\tsnr_2 z_2}\right)\\
r_{\bS_2,\R}>&\tau B\log_2\left(1+\tsnr_2 z_2\right)
\end{cases},
\end{align}
which can further be expressed as
\begin{align}
\begin{cases}\label{case2_2}
z_1\geq&\left(\pra-1\right)(1+\tsnr_2 z_2)/\tsnr_1\\
z_2<&\left(\prb-1\right)/\tsnr_2
\end{cases}.
\end{align}
(\ref{case2_2}) defines the region $\Psi_2$ on the first quadrant of $(z_1,z_2)$ plane, and the probability of Case $2$ is given by

\begin{equation}\label{pm2}
P_{M,2}=\iint\limits_{\Psi_2}p_{z_1,z_2}(z_1,z_2)dz_1dz_2.
\end{equation}
Notice that since the relay cannot decode the message from $\bS_2$ even after interference cancelation, changing the decoding order would not help.
\\
\\
\textbf{Case 3}: This is the symmetric version of Case 2 with the roles of $\bS_1$ and $\bS_2$ interchanged. Hence, the relay can decode the message from $\bS_2$ with interference, but not the message from $\bS_1$. The probability of this case is given by
\begin{equation}\label{pm3}
P_{M,3}=\iint\limits_{\Psi_3}p_{z_1,z_2}(z_1,z_2)dz_1dz_2,
\end{equation}
where $\Psi_3$ is the region in the first quadrant of $(z_1,z_2)$ plane described by
\begin{align}
\begin{cases}\label{case3_2}
z_2\geq&\left(\prb-1\right)(1+\tsnr_1 z_1)/\tsnr_2\\
z_1<&\left(\pra-1\right)/\tsnr_1.
\end{cases}
\end{align}
\\
\\
\textbf{Case 4}: In this case, the relay can decode both messages from two source nodes. Although the description of this case is more involved, we can fortunately express the probability of this case as
\begin{equation}\label{pm4}
P_{M,4}=1-\sum_{i=1}^{3}P_{M,i}.
\end{equation}
Note now that the ON state probability of the $\bS_j-\R$ link is given by
\begin{equation}\label{p12}
P_j=P_{M,j+1}+P_{M,4} \text{ for } j=1,2.
\end{equation}

\subsection{State Probabilities in Broadcast Phase}
In the broadcast phase, the decoding strategy of the destination node $\bD_j$ for $j=1,2$ is described as follows:
\begin{enumerate}
  \item $\bD_j$ attempts to decode its own packet first while treating the interference as noise.
  \begin{enumerate}
    \item If the receiver decodes correctly, then the decoding process ends.
    \item If the receiver cannot decode its own packet first, it tries to decode the packet intended for the other destination first.
    \item If the receiver decodes the other packet correctly, then it moves to the interference cancelation step. Otherwise, it asks for a retransmission from the relay node, and decoding process ceases.
  \end{enumerate}
  \item The receiver performs interference cancelation by subtracting the decoded message from the received signal.
  \item The receiver tries to decode its own packet after interference cancelation. If it still cannot decode the packet, retransmission is required from the relay.
\end{enumerate}

There are two possibilities for link $\R-\bD_1$ being in the ON state. $\bD_1$ may decode its message while treating interference as noise, or it may decode the message for $\bD_2$ first, and then decode its own message after interference cancelation. These are described by the following conditions:
\begin{align}
&r_{\R, \bD_1}\leq(1-\tau)B\log_2\left(1+\frac{\tsnr_r\rho \omega_1}{1+\tsnr_r(1-\rho)\omega_1}\right) \label{eq:broadcast1} \quad \text{ or }\\
&
\begin{cases}\label{case_b11}
r_{\R, \bD_1}&>(1-\tau)B\log_2\left(1+\frac{\tsnr_r\rho \omega_1}{1+\tsnr_r(1-\rho)\omega_1}\right)\\
r_{\R, \bD_1}&\leq(1-\tau)B\log_2(1+\tsnr_r\rho \omega_1)\\
r_{\R, \bD_2}&\leq(1-\tau)B\log_2\left(1+\frac{\tsnr_r(1-\rho)\omega_1}{1+\tsnr_r\rho \omega_1}\right)
\end{cases}
\end{align}
\normalsize
where $\omega_1 = |h_1|^2$. We first define

\begin{align}
 a_1=&\left(\prc-1\right)\bigg/\left\{\tsnr_r\left[1-(1-\rho)\prc\right]\right\}\\
 a_2=&\left(\prc-1\right)/(\tsnr_r\rho)\\
 a_3=&\left(\prd-1\right)\bigg/\left\{\tsnr_r\left[1-\rho\prd\right]\right\}.
\end{align}
\normalsize
Using the conditions in (\ref{eq:broadcast1}) and (\ref{case_b11}), we can express the ON probability of the $\R-\bD_1$ link as

\begin{align}
\hspace{-.2cm} P_3=&
\begin{cases}\label{p3}
0, \hspace{3.4cm} a_1<0 \text{ and } a_3<0\\
\int_{a_1}^{\infty}p_{\omega_1}(\omega_1)d\omega_1, \hspace{0.6cm} (a_1>0 \text{ and }a_3<0)\,\text{or}\,(a_3>a_1>0) \\
\int_{\max\{a_2,a_3\}}^{\infty}p_{\omega_1}(\omega_1)d\omega_1,\hspace{0.3cm} \text{otherwise},
\end{cases}
\end{align}
\normalsize
where, for instance, in Rayleigh fading, $p_{\omega_1}(\omega_1)=\frac{1}{\overline{\omega_1}}\exp(-\omega_1/\overline{\omega_1})$ is the pdf of $\omega_1$, and $\overline{\omega_1}$ is the expected value of $\omega_1$.
A similar analysis can be applied to obtain the ON state probability of the link $\R-\bD_2$ as

\begin{align}
\hspace{-.2cm} P_4=&
\begin{cases}\label{p4}
0, \hspace{3.8cm} b_1<0 \text{ and } b_3<0\\
\int_{b_1}^{\infty}p_{\omega_2}(\omega_2)d\omega_2, \hspace{0.6cm} (b_1>0\text{ and }b_3<0)\,\text{or}\,(b_3>b_1>0)\\
\int_{\max\{b_2,b_3\}}^{\infty}p_{\omega_2}(\omega_2)d\omega_2,\hspace{0.7cm} \text{otherwise},
\end{cases}
\end{align}
\normalsize
where, for instance, if again Rayleigh fading is considered,  $p_{\omega_2}(\omega_2)=\frac{1}{\overline{\omega_2}}\exp(-\omega_2/\overline{\omega_2})$ is the pdf of $\omega_2$, $\overline{\omega_2}$ is the expected value of $\omega_2$, and parameters $b_{j}$ for $j=1,2,3$ are defined as
\begin{small}
\begin{align}
 b_1=&\left(\prd-1\right)\bigg/\left\{\tsnr_r\left[1-\rho\prd\right]\right\}\\
 b_2=&\left(\prd-1\right)/(\tsnr_r(1-\rho))\\
 b_3=&\left(\prc-1\right)\bigg/\left\{\tsnr_r\left[1-(1-\rho)\prc\right]\right\}.
\end{align}
\end{small}
From the view of ARQ, outage happens when the receiver cannot decode the received signal, thus $1-P_j$ can be regarded as outage probabilities of their corresponding links.

\subsection{Stability Conditions} \label{subsec:stability}

Similar to the variable-rate case, stability at the source buffers is ensured by requiring the arrival rates to satisfy (\ref{eq:cond1equ}). The stability at the relay buffer requires the average arrival rate to be smaller than the average departure rate. This can be ensured by choosing the parameters $(\rho,\tau)$ accordingly. Now, our parameter space is just a two dimensional plane, and we can describe the feasible set of $(\rho,\tau)$ for stability as

\begin{equation}\label{stability_1}
\mathbf{\Xi}=\left\{(\rho,\tau)|r_{\bS_1,\R}P_1\leq r_{\R, \bD_1}P_3\:\text{ and }\:r_{\bS_2,\R}P_2\leq r_{\R, \bD_2}P_4\right\}.
\end{equation}
It can be easily seen that both average arrival rates $r_{\bS_1,\R}P_1$ and $r_{\bS_2,\R}P_2$ are monotonic increasing functions of $\tau$, because allocating more time to the multiple-access phase is beneficial to links $\bS_1-\R$ and $\bS_2-\R$. For the same reason, average departure rates $r_{\R, \bD_1}P_3$ and $r_{\R, \bD_2}P_4$ are decreasing functions of $\tau$. Therefore, for given $\rho$, conditions in (\ref{stability_1}) provide two upper bound curves on $\tau$. Then, feasible set for stability is the region under these two upper bounds on the $(\rho,\tau)$ plane.

\subsection{Throughput Region under Statistical Queueing Constraints}
Similar to the variable-rate case, the system throughput is only defined for the feasible parameter setting, which guarantees the stability. For those parameter values outside the stability region, the system throughput is set to $0$. For a given feasible $(\rho,\tau)$ pair and given fixed transmission rates, we next formulate the maximum constant arrival rates $\aR_1$ and $\aR_2$ at the source nodes under statistical queueing constraints parameterized by QoS exponents $\theta_1, \theta_2$ and $\theta_r$. For the described ON-OFF link model with independent fading coefficients, the asymptotic LMGFs can be simplified as
\begin{align}
\Lambda_{\bS_j,\R}(\theta)&=\log\left(e^{\theta r_{\bS_j,\R}}P_j+e^{\theta 0}(1-P_j)\right)\\
                          &=\log\left(e^{\theta r_{\bS_j,\R}}P_j+1-P_j\right) \label{LMGF_s1}\\
    \Lambda_{\R,\bD_j}(\theta)&=\log\left(e^{\theta r_{\R,\bD_j}}P_{j+2}+e^{\theta 0}(1-P_{j+2})\right)\\
                          &=\log\left(e^{\theta r_{\R,\bD_j}}P_{j+2}+1-P_{j+2}\right)\quad j=1,2, \label{LMGF_s2}
%
\end{align}
\normalsize
noting that the transmission rates are either equal to the fixed rates $r_{\bS_j,\R}$ from the sources and $r_{\R,\bD_j}$ from the relay if transmissions are successful and the corresponding links are in the ON state with probabilities $P_j$ and $P_{j+2}$ for $j=1,2$, and are zero in case of failures.
Recall that in order to satisfy the queueing constraints at both the sources and the relay, the arrival rates at the two source nodes should satisfy (\ref{eq:cond1equ}) and (\ref{eq:cond2equ1}) simultaneously. Then, using (\ref{LMGF_s1}) and (\ref{LMGF_s2}) and considering (\ref{eq:cond1equ}), (\ref{eq:cond2equ1}) and (\ref{eq:Lambda_rs2}), we can characterize the maximum constant arrival rates as
\normalsize
{\small
\begin{align}
\aR_1=
\begin{cases}
\min\Big\{-\frac{1}{\theta_1} \log\left(e^{-\theta_1 r_{\bS_1,\R}}P_1+1-P_1\right),
\\
\hspace{0.8cm}-\frac{1}{\theta _{r}}\log \left(e^{-\theta_r r_{\R, \bD_1}}P_3+1-P_3\right)\Big\} &\theta_{r}\le \theta_1 \\
\min\Big\{-\frac{1}{\theta_1} \log\left(e^{-\theta_1 r_{\bS_1,\R}}P_1+1-P_1\right),
\\ \hspace{0.8cm}-\frac{1}{\theta _1}\big(\log \left(e^{-\theta_r r_{\R, \bD_1}}P_3+1-P_3\right)\\
 \hspace{1.2cm}+\log \left(e^{(\theta_r-\theta_1) r_{\bS_1,\R}}P_1+1-P_1\right)\big)\Big\} &\theta_{r}>\theta_1
\end{cases} \label{eq:R1_giventaurho}
\\ 
\aR_2=
\begin{cases}
\min\Big\{-\frac{1}{\theta_2} \log\left(e^{-\theta_2 r_{\bS_2,\R}}P_2+1-P_2\right),
\\
\hspace{0.8cm}-\frac{1}{\theta _{r}}\log \left(e^{-\theta_r r_{\R, \bD_2}}P_4+1-P_4\right)\Big\} &\theta_{r}\le \theta_2 \\
\min\Big\{-\frac{1}{\theta_2} \log\left(e^{-\theta_2 r_{\bS_2,\R}}P_2+1-P_2\right),
\\ \hspace{0.8cm}-\frac{1}{\theta _2}\big(\log \left(e^{-\theta_r r_{\R, \bD_2}}P_4+1-P_4\right)\\
 \hspace{1.2cm}+\log \left(e^{(\theta_r-\theta_2) r_{\bS_2,\R}}P_2+1-P_2\right)\big)\Big\} &\theta_{r}>\theta_2.
\end{cases} \label{eq:R2_giventaurho}
\end{align}}
\normalsize
\hspace{-.1cm}Searching over the stability region $\mathbf{\Xi}$, the arrival rates $\aR_1$, $\aR_2$ and their sum rate can be further optimized over $\rho$ and $\tau$, which will be numerically evaluated in the next subsection.

\subsection{Numerical Results}
\begin{figure}
\begin{center}
\includegraphics[width=0.45\textwidth]{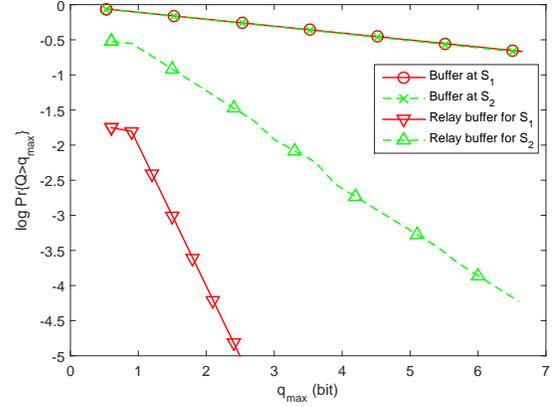}

\caption{Logarithmic buffer overflow probability vs. buffer overflow threshold.}\label{fig_sim3}

\end{center}
\end{figure}

In this subsection, numerical results for the two-source two-destination relay network with fixed transmission rates are provided. First, we verify our analysis through Monte Carlo simulations. In each simulation, we generate $2\times 10^7$ time blocks to estimate the buffer overflow probability, and repeat each simulation $500$ times to evaluate the averages. We set the queueing constraints as $\theta_1=\theta_2=\theta_r=0.1$, and the constant arrival rates at nodes $\bS_1$ and $\bS_2$ are chosen according to (\ref{eq:R1_giventaurho}) and (\ref{eq:R2_giventaurho}), respectively. We further assume that $r_{\bS_1,\R}=r_{\bS_2,\R}=r_{\R, \bD_1}=r_{\R, \bD_2}=0.3\:\text{bit/s}$, $\overline{z_1}=\overline{z_2}=\overline{\omega_1}=\overline{\omega_2}=2$, $\tsnr_1=6.02\:$dB, $\tsnr_2=4.77\:$dB, $\tsnr_r=7.78\:$dB, $\tau=0.39$, $\rho=0.7$. We plot the logarithmic buffer overflow probabilities as functions of the overflow thresholds in Fig. \ref{fig_sim3}. In this specific example, the system throughput is mainly decided by the multiple-access channel, and the overflow probabilities at the buffers of the two source nodes almost exactly meet the queueing constraints. The simulated slopes of the logarithmic overflow probabilities at $\bS_1$ and $\bS_2$ are $-0.099$ and $-0.101$, respectively. The overflow probabilities at the relay buffers diminish with steeper slopes than required and hence satisfy even stricter queueing constraints.  Among the two relay buffers, we note that the overflow probability in the buffer keeping the data from $\bS_1$ decays much more faster. This is because we set $\rho=0.7$, meaning that the $\R-\bD_1$ link gets more power than the $\R-\bD_2$ link.

\begin{figure}
\begin{center}
\includegraphics[width=0.45\textwidth]{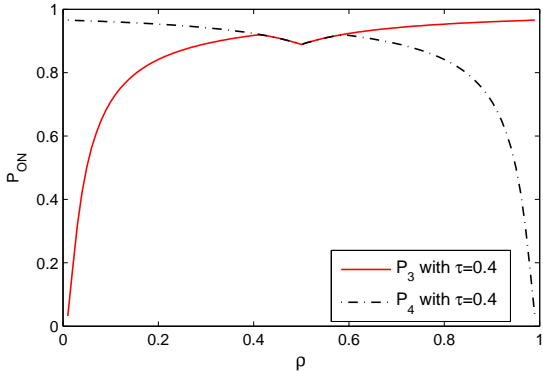}
\caption{ON state probabilities in the broadcast phase vs. $\rho$ with $\tau = 0.4$.}\label{fig2}
\end{center}
\end{figure}

\begin{figure}
\begin{center}
\includegraphics[width=0.45\textwidth]{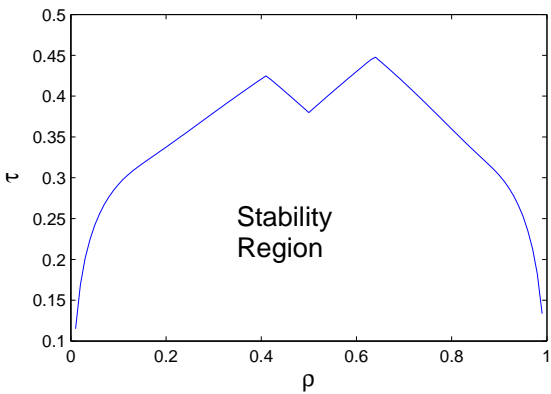}
\caption{Region of feasible $(\rho,\tau)$ pairs for stability at the relay buffer.}\label{fig3}
\end{center}
\end{figure}

The rest numerical results are obtained for the following parameter values: $r_{\bS_1,\R}=r_{\bS_2,\R}=r_{\R, \bD_1}=r_{\R, \bD_2}=0.3\:\text{bit/s}$, $\overline{z_1}=\overline{z_2}=\overline{\omega_1}=\overline{\omega_2}=2$, $\tsnr_1=6.02\:$dB, $\tsnr_2=4.77\:$dB, $\tsnr_r=7.78\:$dB, $\theta_1=\theta_2=1$ and $\theta_r=3$. Fig. \ref{fig2} shows the influence of the power allocation parameter $\rho$ on the successful transmission probabilities $P_3$ and $P_4$ in the broadcast phase. We observe that as $\rho$ increases from $0$ to $0.5$ and hence a larger fraction of the power is allocated to the transmission of the message to $\bD_1$, $P_3$ grows dramatically while $P_4$ diminishes by a relatively small amount. This indicates that the sum arrival rate increases initially with increasing $\rho$. We also notice that both $P_3$ and $P_4$ decrease slightly at around $\rho=0.5$ due to the increased interference caused by the joint transmission of messages at similar power levels in the broadcast phase. Similarly, the boundary of region of feasible  $(\rho,\tau)$ pairs for stability at the relay buffer shown in Fig. \ref{fig3} has a local minimum $\tau$ value at $\rho$ close to $0.5$. Additionally, we see in the figure that the boundary, which essentially bounds $\tau$ from above, can be regarded as the intersection of two upper bounds on $\tau$ as discussed in Section \ref{subsec:stability}.

\begin{figure}

\begin{center}
\includegraphics[width=0.45\textwidth]{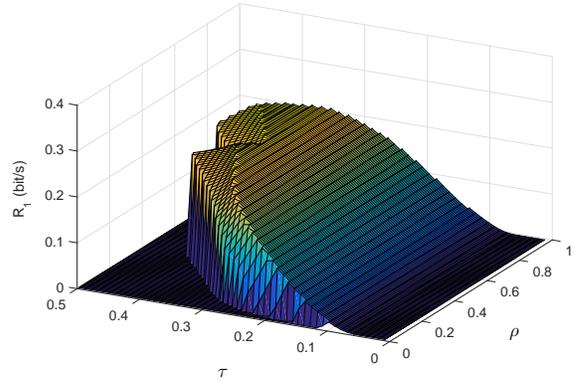}

\caption{The arrival rate $\aR_1$ as a function of $(\rho,\tau)$.}\label{fig4}

\end{center}
\end{figure}

Fig. \ref{fig4} shows the arrival rate $\aR_1$ as a function of $(\rho,\tau)$. Outside the feasible region, rate is set to zero. We note that as $\tau$ increases, $\aR_1$ initially increases and then decreases within the feasible region. From (\ref{eq:R1_giventaurho}), we know that $\aR_1$ is characterized as the point-wise minimum of two functions, one being an increasing function of $\tau$, while the other being a decreasing function. Therefore, there exists an optimal $\tau$ that maximizes $\aR_1$ for given $\rho$. We also observe that the maximum of $\aR_1$ over all $(\rho,\tau)$ is achieved when $\tau = 0.39$ and $\rho = 0.7$. Note that with this relatively large $\rho$ value,  the $\bS_1-\R-\bD_1$ link can in general support higher arrival rates because the relay allocates more power for the transmission of the message coming from $\bS_1$. Similar numerical results can be obtained for $\aR_2$. Expectedly, $\aR_2$ has higher values when $\rho <0.5$.

\begin{figure}
\begin{center}
\includegraphics[width=0.45\textwidth]{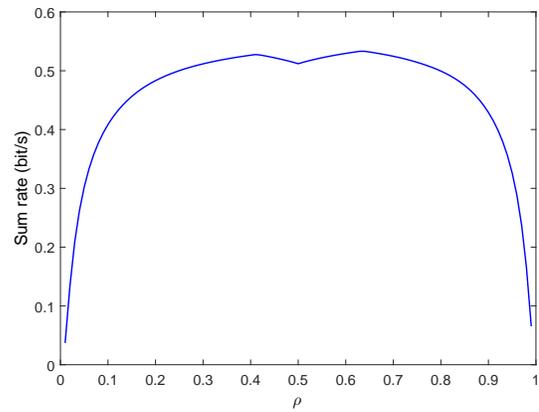}

\caption{Maximum sum arrival rate vs. $\rho$.}\label{fig5}

\end{center}
\end{figure}

Finally, we consider the maximum sum arrival rate. Fig. \ref{fig5} plots the maximum sum arrival rate $\max\{\aR_1+\aR_2\}$ as a function of $\rho$. As $\rho$ approaches $0$, the performance of the $\bS_1-\R-\bD_1$ link is limited by the low transmission power of the relay, leading to the adoption of a very small $\tau$ value as seen in Fig. \ref{fig3}. Small value of $\tau$ lowers the throughput of the $\bS_2-\R-\bD_2$ link as well. Hence $\aR_2$ is also small. Similar concerns arise as $\rho$ approaches $1$. Hence, allocating the power almost exclusively for the transmission of one message is not an efficient strategy in terms of maximizing the sum rate. Indeed, the sum arrival rate is maximized when $\rho = 0.64$. However, it is interesting to note that equal allocation (i.e., having $\rho = 0.5$) is not the optimal strategy either, because the sum rate has a local minimum point around $\rho=0.5$ again as a reflection of increased interference.


\section{Conclusion}

In this paper, we have studied the throughput of multi-source multi-destination relay networks under statistical queueing constraints, for both cases of with and without CSI at the transmitter sides. When there is perfect CSI at the transmitter, transmission rates can be varied according to the instantaneous channel conditions. We have characterized the instantaneous channel capacities in different phases as functions of the system parameters $\tau$, $\rho$ and $\delta$. When CSI is not available at the transmitter side, transmissions are performed at fixed rates, and decoding failures lead to retransmission requests via an ARQ protocol. We have modeled the links to be in ON or OFF states depending on the reliability of the reception. We have determined the probabilities of these states.

Following these characterizations, we have described, for both perfect and no CSI cases, the stability conditions, and defined the feasible region of the transmission parameters. For the variable-rate scheme, the stability region is defined in the three dimensional space of the system parameters $\tau$, $\rho$, and $\delta$. For the fixed rate scheme, the stability region is defined on the $\rho-\tau$ plane to ensure stability at the relay buffers.

Finally, we have characterized the arrival rates under queueing constraints at the source and relay nodes as a function of the QoS exponents, channel fading and system parameters for both cases. In addition, the concavity of the throughput function is shown with respect to the system parameters $\delta$ and $\tau$ for the variable-rate scheme. Also in the variable-rate model, two extensions have been addressed, namely the multi-source multi-destination relay network with more than two source-destination pairs and full-duplex relaying. We have verified the theoretical results via Monte Carlo simulations. Numerically, for the variable-rate model, we have investigated the optimal position of the relay node. Also, the throughput region is obtained via searching over the three dimensional parameter space. For the fixed rate scheme, we have determined the feasible region of $(\rho,\tau)$ and investigated the impact of these transmission parameters on the arrival rates. We have noted that sum arrival rate is maximized when relay allocates comparable power levels to the transmission of the messages of $\bS_1$ and $\bS_2$ while increased interference can slightly diminish the performance when $\rho$ is very close to 0.5.

\bibliographystyle{ieeetr}
\bibliography{Relay_ARQ}

\end{document}